\documentclass[aps, prl, twocolumn, superscriptaddress, amsmath 
]{revtex4-2}

\usepackage{graphicx}
\usepackage{amsfonts}
\usepackage{amssymb}
\usepackage{amsmath}
\usepackage{lineno}
\usepackage[normalem]{ulem}
\usepackage{hyperref}
\hypersetup{
    colorlinks=true,       
    linkcolor=blue,          
    citecolor=blue,        
    filecolor=blue,      
    urlcolor=blue           
}
\usepackage{makecell}
\usepackage{tabularx}
\usepackage{array}

\newcommand{\ket}[1]{\vert#1\rangle}
\newcommand{\bra}[1]{\langle #1\vert}

\newcommand{\be}{\begin{equation}}
\newcommand{\ee}{\end{equation}}
\newcommand{\ba}{\begin{array}}
\newcommand{\ea}{\end{array}}
\newcommand{\bea}{\begin{eqnarray}}
\newcommand{\eea}{\end{eqnarray}}

\def\6{{\langle}}
\def\9{{\rangle}}

\usepackage{xcolor}

\begin{document}

\title{Entangling independent particles by path identity}

\author{Kai Wang}
\email{These two authors contributed to this work equally}
\affiliation{National Laboratory of Solid-state Microstructures, School of Physics, Collaborative Innovation Center of Advanced Microstructures, Nanjing University, Nanjing 210093, China}
	
\author{Zhaohua Hou}
\email{These two authors contributed to this work equally}
\affiliation{National Laboratory of Solid-state Microstructures, School of Physics, Collaborative Innovation Center of Advanced Microstructures, Nanjing University, Nanjing 210093, China}
	
\author{Kaiyi Qian}
\affiliation{National Laboratory of Solid-state Microstructures, School of Physics, Collaborative Innovation Center of Advanced Microstructures, Nanjing University, Nanjing 210093, China}

\author{Leizhen Chen}
\affiliation{National Laboratory of Solid-state Microstructures, School of Physics, Collaborative Innovation Center of Advanced Microstructures, Nanjing University, Nanjing 210093, China}

\author{Mario Krenn}
\email{Corresponding author: mario.krenn@mpl.mpg.de}
\affiliation{Max Planck Institute for the Science of Light (MPL), Erlangen, Germany}
	
\author{Shining Zhu}
\affiliation{National Laboratory of Solid-state Microstructures, School of Physics, Collaborative Innovation Center of Advanced Microstructures, Nanjing University, Nanjing 210093, China}
	
\author{Xiao-Song Ma}
\email{Corresponding author: Xiaosong.Ma@nju.edu.cn}
\affiliation{National Laboratory of Solid-state Microstructures, School of Physics, Collaborative Innovation Center of Advanced Microstructures, Nanjing University, Nanjing 210093, China}
	
\date{\today}
	
\begin{abstract}
Quantum entanglement — correlations of particles that are stronger than any classical analogue — is the basis for research on the foundations of quantum mechanics and for practical applications such as quantum networks. Traditionally, entanglement is achieved through local interactions or via entanglement swapping, where entanglement at a distance is generated through previously established entanglement and Bell-state measurements. However, the precise requirements enabling the generation of quantum entanglement without traditional local interactions remain less explored. Here we demonstrate that independent particles can be entangled without the need for direct interaction, prior established entanglement, or Bell-state measurements, by exploiting the indistinguishability of the origins of photon pairs. Our demonstrations challenge the long-standing belief that the prior generation and measurement of entanglement are necessary prerequisites for generating entanglement between independent particles that do not share a common past. In addition to its foundational interest, we show that this technique might lower the resource requirements in quantum networks, by reducing the complexity of photon sources and the overhead photon numbers.
\end{abstract}

\maketitle

Quantum entanglement is usually created by a local interaction between particles \cite{PhysRev.47.777,PhysRev.48.696,Schrödinger1935}, for example via scattered annihilation radiation \cite{PhysRev.77.136}, atomic cascades \cite{PhysRevLett.28.938,PhysRevLett.49.1804}, nonlinear optical effects \cite{PhysRevLett.61.2921,PhysRevLett.61.50,PhysRevLett.71.3893,HARDY1992326,PhysRevLett.75.4337,PhysRevLett.115.250401,PhysRevLett.115.250402}, and other types of interactions between light / microwaves / phonons and matters \cite{PhysRevLett.79.1,Julsgaard2001,Majer2007,Jost2009,Riedinger2016,Ockeloen-Korppi2018,Kurpiers2018,Bienfait2019,PhysRevLett.125.260502,Storz2023}. However, entanglement does not always have to be generated at the same location. In 1993, it was theoretically predicted that spatially separated photons could be entangled in an ‘entanglement swapping’ procedure \cite{PhysRevLett.70.1895,PhysRevLett.71.4287} (see Fig. 1(a)). In this method, two pairs of entangled photons are used. One photon from each pair is sent to a measurement station, where these two photons are projected into a maximally entangled Bell state. This procedure is typically referred to as a ‘Bell-state measurement’ (BSM). This joint measurement causes the other two photons — which have neither interacted with each other nor shared any common past — to become entangled.

Observed for the first time by Pan et al. in 1998 \cite{PhysRevLett.80.3891}, entanglement swapping soon became an essential building block for quantum-communication networks \cite{PhysRevLett.81.5932,Duan2001,PhysRevLett.90.207901,RevModPhys.83.33,RevModPhys.95.045006} and the conceptual basis for experiments at the foundations of quantum mechanics \cite{Ma2012,Hensen2015,PhysRevLett.119.010402}. Entanglement swapping requires preparing entangled states in advance and then performing the BSM on the two ancillary photons, which relies on Hong–Ou–Mandel interference of two photons \cite{PhysRevLett.59.2044,PhysRevA.67.022301,PhysRevLett.96.110501,PhysRevLett.96.240502,PhysRevA.79.040302}.

Here we demonstrate a fundamentally different way to entangle two independent particles, without any direct interaction or sharing any common past. Our method is inspired by path identity in nonlinear interferometers, also known as frustrated interference, which was first discovered by Zou, Wang, and Mandel \cite{PhysRevLett.67.318} and later by Herzog et al. \cite{PhysRevLett.72.629}, respectively. See ref \cite{10.1063/1.881360,RevModPhys.94.025007} for reviews on this topic. In frustrated interference, the photon-pair-generation processes can be suppressed or promoted by varying the interferometric phases. The specific scheme of our work is discovered using PyTheus, which is an efficient, automated design and discovery framework for quantum-optics experiments \cite{RuizGonzalez2023digitaldiscoveryof}. By employing the four-photon frustrated interference \cite{Gu2019,Feng:23,Qian2023}, we can create entangled pairs of spatially separated photons, (1) without the creation of prior entanglement, (2) without a Bell-state measurement, and (3) even without measuring both of the ancillary photons. Each of these three points is necessary for entanglement swapping, demonstrating that our work is a complementary mechanism of a key protocol for quantum information processing with less resource requirements.

\begin{figure*}
\includegraphics[width=16cm]{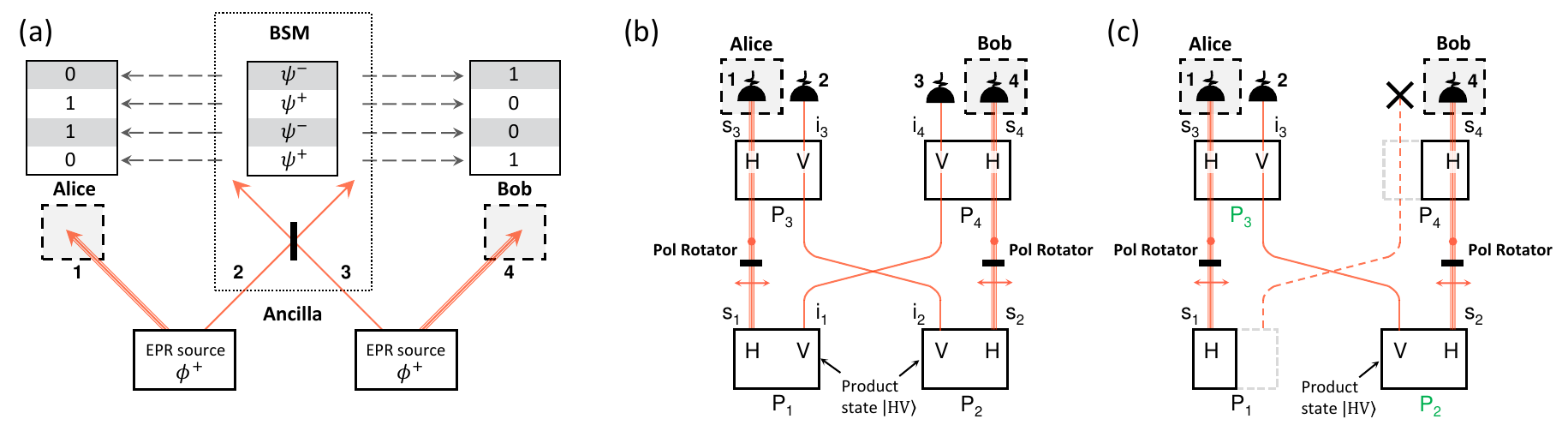}
\caption{Schemes for entangling independent photons. (a) Entanglement swapping. Two Einstein–Podolsky–Rosen (EPR) sources produce two pairs of photons in the state $\ket{\phi^+}=\frac{1}{\sqrt{2}}(\ket{\rm HH}+\ket{\rm VV})$. Ancillary photons on paths 2 and 3 are sent to the central Bell-state measurement (BSM) station and are projected onto one of the Bell states. Upon this process, Alice and Bob share a maximum entangled state between photons on paths 1 and 4. (b) Entanglement generation by path identity. Every white square ($\rm P_1$–$\rm P_4$) represents a probabilistic two-photon source that can generate the polarization product state, $\ket{\rm HV}$. The paths of photons possibly produced by sources $\rm P_1$ and $\rm P_2$ (lower row) are overlapped with the paths of photons that can be produced by $\rm P_3$ and $\rm P_4$ (upper row). Photons from sources $\rm P_1$ and $\rm P_2$ on paths 1 ($\rm s_1$) and 4 ($\rm s_2$) each pass through a polarization rotator, which rotates their polarization states from horizontal to vertical. As a consequence, photons in the same path are indistinguishable except for their polarization. In this configuration, Alice and Bob share a maximally entangled state between photons on paths 1 and 4, conditioned on the coincidence of photons on paths 2 and 3 in a polarization product state. (c) Entanglement generation by path identity while detecting only one ancillary photon 2. When we reduced the pump intensities of $\rm P_2$ and $\rm P_3$ to suppress the generation probability of sources $\rm P_2$ ($\rm i_2$, $\rm s_2$) and $\rm P_3$ ($\rm i_3$, $\rm s_3$), we generate entanglement between Alice (photon on path 1) and Bob (photon on path 4) by detecting only photons on paths 1, 2, and 4. This scheme can also be interpreted as an experiment that only has two photon pair sources and two single-photon sources ($\rm s_1$ and $\rm s_4$).}
\end{figure*}

\section{Generate entanglement between non-interacting photons}
The scheme for the first level of our demonstration is shown in Fig. 1(b). We have four photon-pair sources ($\rm P_1$–$\rm P_4$) based on Spontaneous Parametric Down-Conversion (SPDC). Each can create a polarization product state $\ket{\rm HV}$ in a probabilistic way. We detect four photons in our experiment. The photons from sources $\rm P_1$ and $\rm P_2$ are aligned with those from sources $\rm P_3$ and $\rm P_4$ to ensure that their paths are indistinguishable. On paths 1 and 4 (thick red lines), we introduce polarization rotators, which transform the horizontally polarized signal photons ($\rm s_1$, $\rm s_2$) from sources $\rm P_1$ and $\rm P_2$ to vertically polarized ones. The idler photons on paths 2 ($\rm i_2$, $\rm i_3$) and 3 ($\rm i_1$, $\rm i_4$) (thin red lines) are always vertically polarized and are indistinguishable in every degrees of freedom. Conditioned on the detection of photons in paths 2 and 3, the photons of Alice (path 1) and Bob (path 4) are entangled. The final output state of the photons on the four paths 1–4 is (to the second-order approximation of the SPDC process):

\begin{align}
\nonumber \ket{\psi_f}&=\varepsilon_1\varepsilon_2\ket{\rm VVVV}_{1234}+\varepsilon_3\varepsilon_4\ket{\rm HVVH}_{1234}\\
\nonumber &+\varepsilon_1\varepsilon_3\ket{\rm HV}_{1}\ket{\rm VV}_{23}+\sqrt{2}\varepsilon_1\varepsilon_4\ket{\rm VH}_{14}\ket{\rm V^2}_3\\
&+\sqrt{2}\varepsilon_2\varepsilon_3\ket{\rm HV}_{14}\ket{\rm V^2}_2+\varepsilon_2\varepsilon_4\ket{\rm HV}_{4}\ket{\rm VV}_{23},
\end{align}
where $\varepsilon_i$ (i = 1, 2, …, 4) represents the generation efficiency of source $\rm P_i$ (see Appendix A1 for details). The subscript represents the path number. $\ket{\rm V^2}_i$ stands for two vertically polarized photons on path i, and $\ket{\rm HV}_{i}$ represents the situation with both horizontally and vertically polarized photons on path i. We set all the sources to the same generation efficiency $\varepsilon$. In this process, all the two-photon sources provide the possibility to generate a polarization product state. We then detect four-photon coincidence events (i.e., one photon on each path). This corresponds to the projection of the state shown in the first line of Eq. 1, and we thereby obtain the following state:
\begin{equation}
\ket{\varphi}=\varepsilon^2(\ket{\rm HH}+\ket{\rm VV})_{14}\ket{\rm VV}_{23}.
\end{equation}

This is a maximally entangled state of photons on paths 1 and 4, conditioned on detecting a polarization product state in paths 2 and 3. Therefore, unlike entanglement swapping, we need neither to prepare entanglement in advance, nor to rely on two-photon interference to realize a BSM. 

Moreover, another unique feature of our scheme compared to entanglement swapping is that we even do not need to detect all the ancillary photons to create entanglement. As shown in Fig. 1(c), in the second level of our demonstration, we detect only photons 1, 2, and 4, and obtain the following three terms from Eq. 1:
\begin{align}
\nonumber \ket{\varphi'}&=(\varepsilon_3\varepsilon_4\ket{\rm HH}+\varepsilon_1\varepsilon_2\ket{\rm VV})_{14}\ket{\rm VV}_{23}\\
&+\sqrt{2}\varepsilon_2\varepsilon_3\ket{\rm HV}_{14}\ket{V^2}_{2},
\end{align}

\begin{figure*}
\includegraphics[width=16cm]{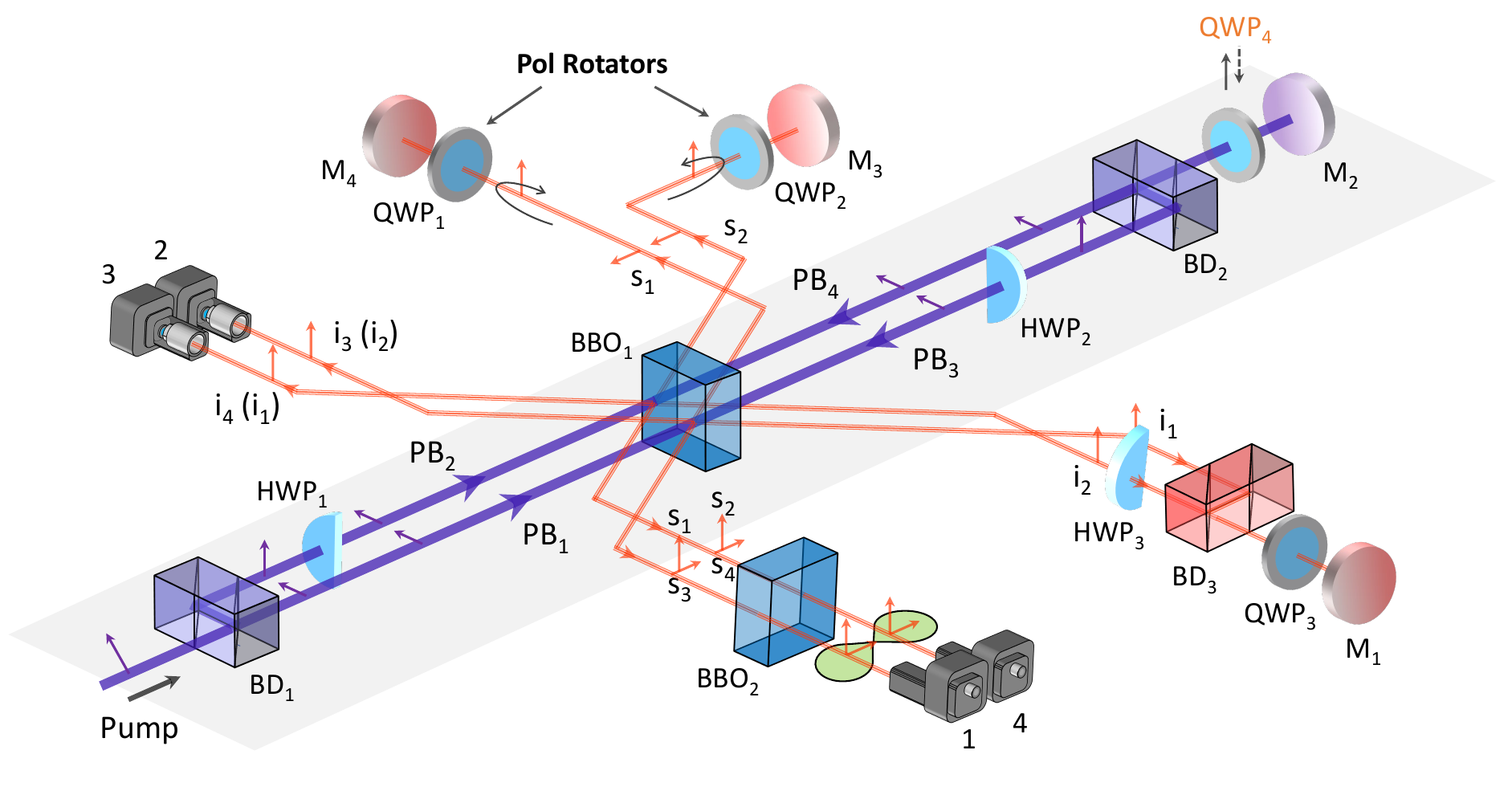}
\caption{Experimental setup. We use two parallel reflect-type frustrated interferometer to probabilistically generate two pairs of photons ($\rm i_1$, $\rm s_1$) \& ($\rm i_2$, $\rm s_2$), or ($\rm i_3$, $\rm s_3$) \& ($\rm i_4$, $\rm s_4$) from a nonlinear crystal ($\rm BBO_1$). The idler photons $\rm i_1$ and $\rm i_2$ swap their path by polarization and are reflected back by mirror $\rm M_1$. A quarter-wave plate ($\rm QWP_4$) before $\rm M_2$ swaps the two pumps $\rm PB_1$ and $\rm PB_2$, and is removed in the first level of our demonstration (see Fig. 1(b)). The horizontally polarized signal photons on paths 1 ($\rm s_1$)/4 ($\rm s_2$) are rotated to vertical polarization by the polarization rotators ($\rm QWP_1$ and $\rm QWP_2$) and reflected by $\rm M_4$/$\rm M_3$. Then they are superimposed with the photons ($\rm s_3$, $\rm s_4$) generated by the reflected pump ($\rm PB_3$, $\rm PB_4$). The temporal difference of photons on the same path can be eliminated by tuning the position of the reflectors ($\rm M_1$-$\rm M_4$) and the spatial walk-off is compensated by $\rm BBO_2$. We generate entanglement (green circles) between Alice (path 1) and Bob (path 4) in the state $\ket{\psi}_{14}=\frac{1}{\sqrt{2}}(\ket{\rm HH}+\ket{\rm VV})$, conditioned on the simultaneous detection of photons on paths 2 and 3 in a polarization product state. We further reduce the intensity of pumps $\rm PB_2$ and $\rm PB_3$ compared to $\rm PB_1$ and $\rm PB_4$ with $\rm QWP_4$, and generate entanglement between Alice and Bob by only detecting photons on path 2. This is the second level of our demonstration (see Fig. 1(c)). All photons are collected with single-mode fiber couplers (black assemblies) and detected with single-photon detectors. (HWP, half-wave plate; BD, beam displacer.)}
\end{figure*}

When we suitably reduce the generation efficiencies of sources $\rm P_2$ ($\varepsilon_2$) and $\rm P_3$ ($\varepsilon_3$) (shown in green letters in Fig. 1(c) indicating power reductions), and increase those of sources $\rm P_1$ ($\varepsilon_1$) and $\rm P_4$ ($\varepsilon_4$), the third term in Eq. 3 can be made arbitrarily small, and hence we can generate entanglement between photons 1 and 4, conditioned only on detecting photon 2. One interesting interpretation of Fig. 1(c) is the following: As the photon in path c is never detected, one could argue it does not need to be generated in the first place. Therefore, one can reinterpret the setup as one that combines two-photon pair sources ($\rm P_2$ and $\rm P_3$) and two single-photon emitters ($\rm P_1$ and $\rm P_4$, emitting into path 1 and 4, respectively). It requires the coherent superposition of single-photon and photon-pair sources, which is experimentally possible, as shown in a remarkable demonstration \cite{PhysRevLett.87.123603}.

\section{Experimental Setup}

Our experimental scheme is shown in Fig. 2. The pump are femtosecond pulses with the central wavelength at 404 nm (80 MHz repetition rate) with diagonal polarization. It enters from the left side of the setup and splits on a beam displacer ($\rm BD_1$). We use the two parallel pump beams $\rm PB_1$ and $\rm PB_2$ to pump one $\beta$-barium borate crystal ($\rm BBO_1$) and probabilistically generate two pairs of photons in the state $\ket{\psi}_{\rm s_1 s_2 i_1 i_2}=\ket{\rm HHVV}$ in the beam-like configuration \cite{Takeuchi:01,Niu:08}. Here we use $\rm P_i$ to denote the source and $\rm PB_i$ to denote the pump beam used to pump the respective source. The signal photons $\rm s_1$ and $\rm s_2$ are reflected back from mirrors $\rm M_4$ and $\rm M_3$, respectively. As $\rm s_1$ and $\rm s_2$ pass through their respective polarization rotators (quarter-wave plates $\rm QWP_1$ and $\rm QWP_2$, fixed at $45^{\circ}$) twice, their polarizations are rotated to the vertical direction on the way back. The idler photons $\rm i_1$ and $\rm i_2$ are combined on $\rm BD_3$ and reflected on $\rm M_1$. They pass through $\rm QWP_3$ twice, so that we can swap their path using their polarizations.

$\rm PB_1$ and $\rm PB_2$ are then reflected by $\rm M_2$ and pump the BBO crystal again ($\rm PB_3$, $\rm PB_4$). They probabilistically generate the photons in the state $\ket{\psi}_{\rm s_3 s_4 i_3 i_4}=\ket{\rm HHVV}$. To generate entanglement, we erase the temporal distinguishability of the photons on the same path. By tuning $\rm M_1$, we change the time delay of photons $\rm i_1$ and $\rm i_2$. Thereby, we ensure that they arrive simultaneously with the pump $\rm PB_3$ and $\rm PB_4$, respectively, at $\rm BBO_1$. For the photons $\rm s_1$ and $\rm s_2$, we move $\rm M_4$ and $\rm M_3$ to compensate for the temporal difference between $\rm s_1$ \& $\rm PB_3$, and $\rm s_2$ \& $\rm PB_4$. However, their vertical polarization leads to an additional spatial walk-off on $\rm BBO_1$ compared to $\rm s_3$ and $\rm s_4$ due to birefringence. Therefore, we use $\rm BBO_2$ which has the same thickness as $\rm BBO_1$ but with an optical axis that rotates $180^{\circ}$ with respect to $\rm BBO_1$ to compensate for the spatial walk-off. Then we have the coherent superposed state $\ket{\psi}_{1234}=\frac{1}{\sqrt{2}}(\ket{\rm VVVV}+\ket{\rm HVVH})$. By post-selecting four-photon detection events, we generate entanglement between Alice (photon 1) and Bob (photon 4): $\ket{\psi}_{14}=\frac{1}{\sqrt{2}}(\ket{\rm VV}+\ket{\rm HH})$ conditioned on the simultaneous detection of $\ket{\rm VV}$ of photons 2 and 3. Moreover, we adjust the power of pumps $\rm PB_2$ and $\rm PB_3$, and entangle photons 1 and 4 by detecting only photon 2 (see Eq. 3). The four photons are collected by single-mode fiber couplers 1-4 (black assemblies in Fig. 2). Later, we detect photons 2 and 3 (or photon 2 alone) directly using single-photon detectors. Alice and Bob perform the respective polarization measurements on photons 1 and 4 in order to verify the entanglement between them. 

\begin{figure}
\includegraphics[width=8.5cm]{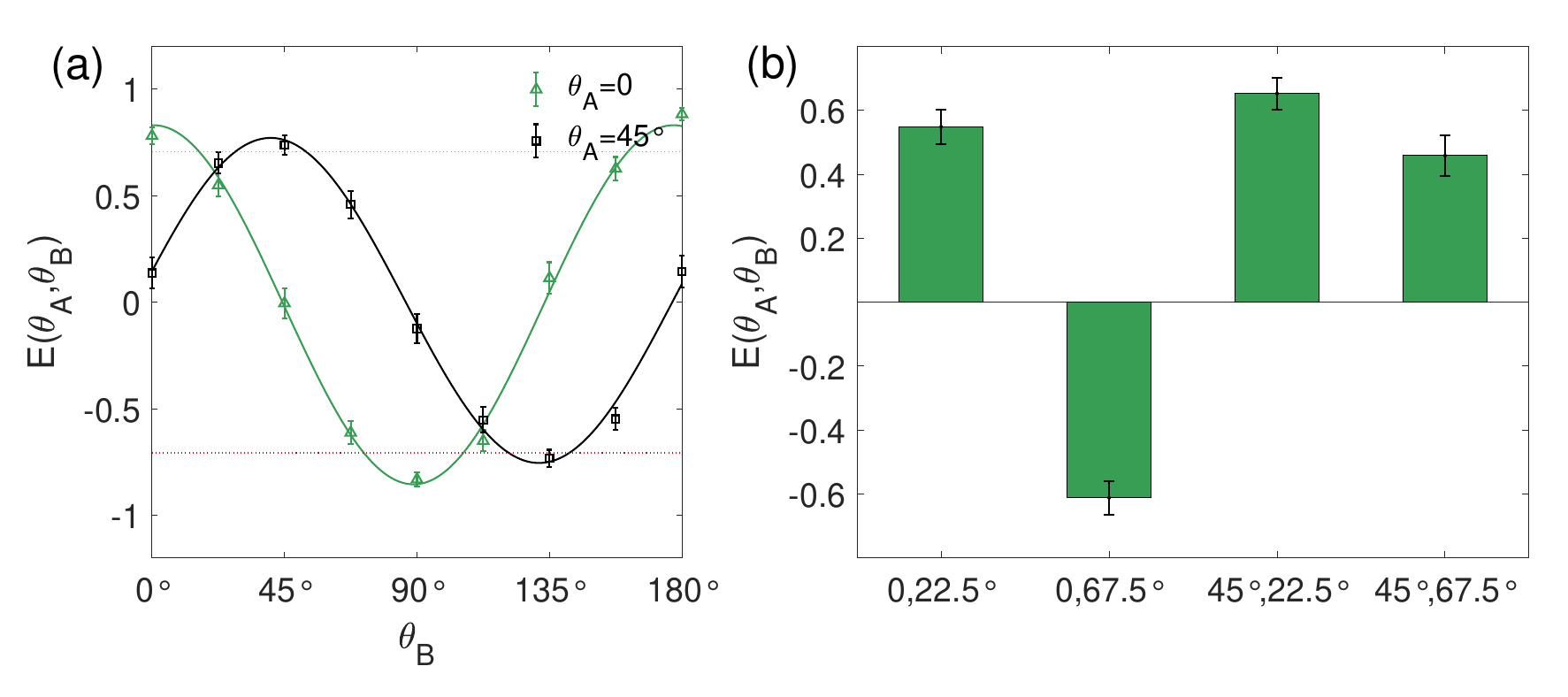}
\caption{Experimental correlation functions of photons 1 and 4 conditioned on the simultaneous detection of photons 2 and 3 in a polarization product state, $\ket{\rm VV}_{23}$. (a) The correlation function of Alice and Bob. Alice sets $\theta_{A}$ to $0^{\circ}$ (green data points, fitted with the green curve) and $45^{\circ}$ (black data points, fitted with the black curve), while Bob sweeps $\theta_{B}$ from $0^{\circ}$ to $180^{\circ}$. (b) The four expectation values $E(0^{\circ}, 22.5^{\circ})$, $E(0^{\circ}, 67.5^{\circ})$, $E(45^{\circ}, 22.5^{\circ})$, $E(45^{\circ}, 67.5^{\circ})$ used to calculate the S parameter of the CHSH inequality: $E(0^{\circ}, 22.5^{\circ}) = 0.5490 \pm 0.0541$, $E(0^{\circ}, 67.5^{\circ}) = - 0.6121 \pm 0.0531$, $E(45^{\circ}, 22.5^{\circ}) = 0.6528 \pm 0.0497$, $E(45^{\circ}, 67.5^{\circ}) = 0.4586 \pm 0.0642$. Error bars are derived from Poissonian statistics and error propagation.}
\end{figure}

\section{Results}
To verify the entanglement of the generated quantum state $\ket{\psi}_{14}$ distributed to Alice and Bob, we firstly demonstrate the violation of the Clauser–Horne–Shimony–Holt (CHSH) version of Bell’s inequality \cite{PhysRevLett.23.880} with state $\ket{\psi}_{14}$. We measure the normalized expectation value based on the results of polarization measurements performed by Alice (polarization analyzer at the angle of $\theta_{A}$) and Bob ($\theta_{B}$), $E(\theta_{A}, \theta_{B})$, which is defined as: 

\begin{equation}
E(\theta _A,\theta _B)=\frac{N_{++}-N_{+-}-N_{-+}+N_{--}}{N_{++}+N_{+-}+N_{-+}+N_{--}}
\end{equation}
where $N_{ij}$ represents the coincidence counts of the outcome $i$, $j$ (+ is for parallel, – for antiparallel of the angle of two analyzers) at the measurement setting $\theta_{A}$, $\theta_{B}$. The experimental results of these correlation measurements are shown in Fig. 3(a), in which Alice fixes her polarization analyzer at the angle $\theta_{A}=0^{\circ}$ (green data points, fitted with green line) and $45^{\circ}$ (black data points, fitted with black line), respectively. Bob sweeps the angle of his polarization analyzer, $\theta_{B}$, from $0^{\circ}$ to $180^{\circ}$ to measure the correlation between photons 1 and 4. We denote $0^{\circ}$ as the horizontal polarization and $90^{\circ}$ as the vertical polarization. The maximum values of the correlation functions are $0.8827\pm0.0294$ for $\theta_{A}=0^{\circ}$, and $0.7379\pm0.0466$ for $\theta_{A}=45^{\circ}$, respectively. Both are above the classical bound $1/\sqrt{2}$ required for violating CHSH inequality, showing the nonlocality of the generated entangled state. Our results are shown in Fig. 3(b). Based on these correlation functions $E(\theta_{A}, \theta_{B})$, we obtain the S-value of the CHSH inequality:$S=|E(0^{\circ},22.5^{\circ})-E(0^{\circ},67.5^{\circ})+E(45^{\circ},22.5^{\circ})+E(45^{\circ},67.5^{\circ})|=2.2724\pm 0.0822$, which violates the inequality by more than three standard deviations. For complete data of joint probability distributions of photons 1 and 4 in different bases, see Appendix A2.

\begin{figure}
\includegraphics[width=8.5cm]{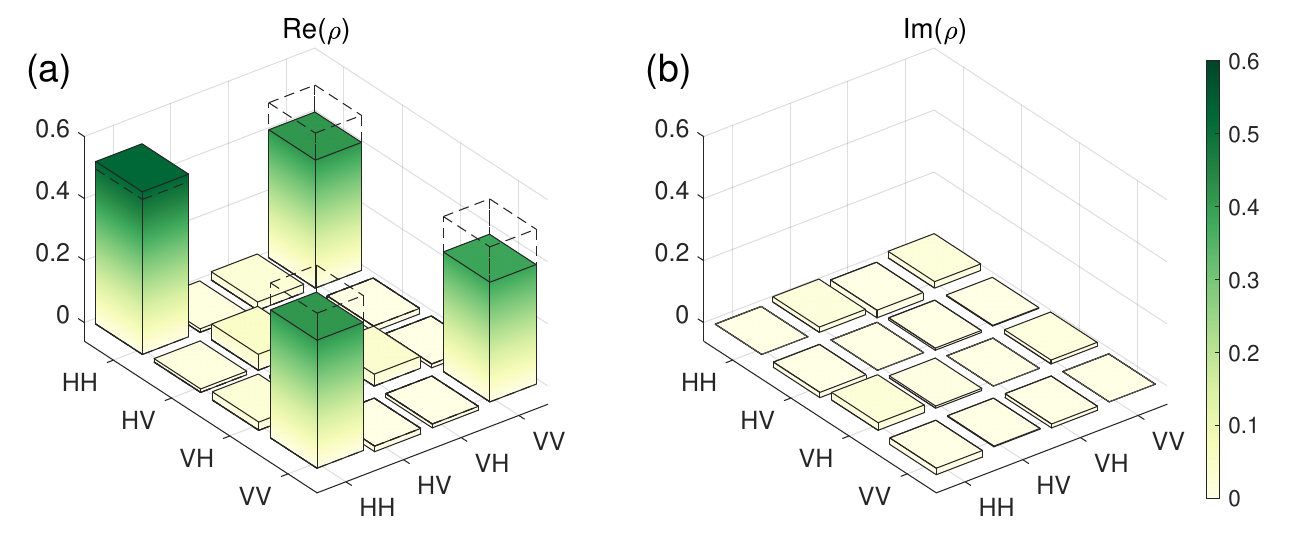}
\caption{Quantum-state-tomography results conditioned on the simultaneous detection of photons 2 and 3 in $\ket{\rm VV}_{23}$. (a) Real and (b) the imaginary parts of the density matrix of state $\ket{\psi}_{14}$. The experimental quantum state has a fidelity of $0.868\pm0.007$ with respect to the Bell state $\ket{\phi^+}=\frac{1}{\sqrt{2}}(\ket{\rm HH}+\ket{\rm VV})$. The dashed wire grid represents the ideal case for state $\ket{\phi^+}$. Uncertainties are obtained from 100 Monte Carlo simulations with counting Poisson statistics.}
\end{figure}

We further perform quantum state tomography to completely characterize the entangled state of Alice and Bob. The result is shown in Fig. 4. The fidelity of state $\ket{\psi}_{14}$ is $F=0.868\pm0.007$ with respect to the Bell state $\ket{\phi^+}=\frac{1}{\sqrt{2}}(\ket{\rm HH}+\ket{\rm VV})$. The concurrence of the experimentally obtained state is $0.746\pm0.013$, unambiguously showing the existence of quantum entanglement between photons 1 and 4 shared by Alice and Bob. We present the method to determine the relative phase of the entangled states between photons 1 and 4, see Appendix A3.

Next we introduce the second-level demonstration of our work to create remote entanglement by detecting only one ancillary photon. As shown schematically in Fig. 1(c), and mathematically in Eq. 3, we can detect only one ancillary photon, photon 2, to create entanglement between Alice and Bob. To achieve this, we reduce the pump intensity of sources $\rm P_2$ and $\rm P_3$. Here we set one QWP ($\rm QWP_4$ in Fig. 2) with its optical axis at $45^{\circ}$ on the path of the pump after $\rm BD_2$. Therefore, pumps $\rm PB_1$ and $\rm PB_2$ are swapped on their way back. $\rm PB_2$ is reflected as $\rm PB_3$, with the same intensity. We rotate the polarization of the incident pump before $\rm BD_1$ to tune the intensity of pump $\rm PB_1$ ($\rm PB_4$) and pump $\rm PB_2$ ($\rm PB_3$). Here $\varepsilon_1 = \varepsilon_4 = \varepsilon$ and $\varepsilon_2 = \varepsilon_3 = \varepsilon'$, and the ratio $\frac{\varepsilon'}{\varepsilon}\approx 0.184$ (for details see Appendix A4).

\begin{figure}
\includegraphics[width=8.5cm]{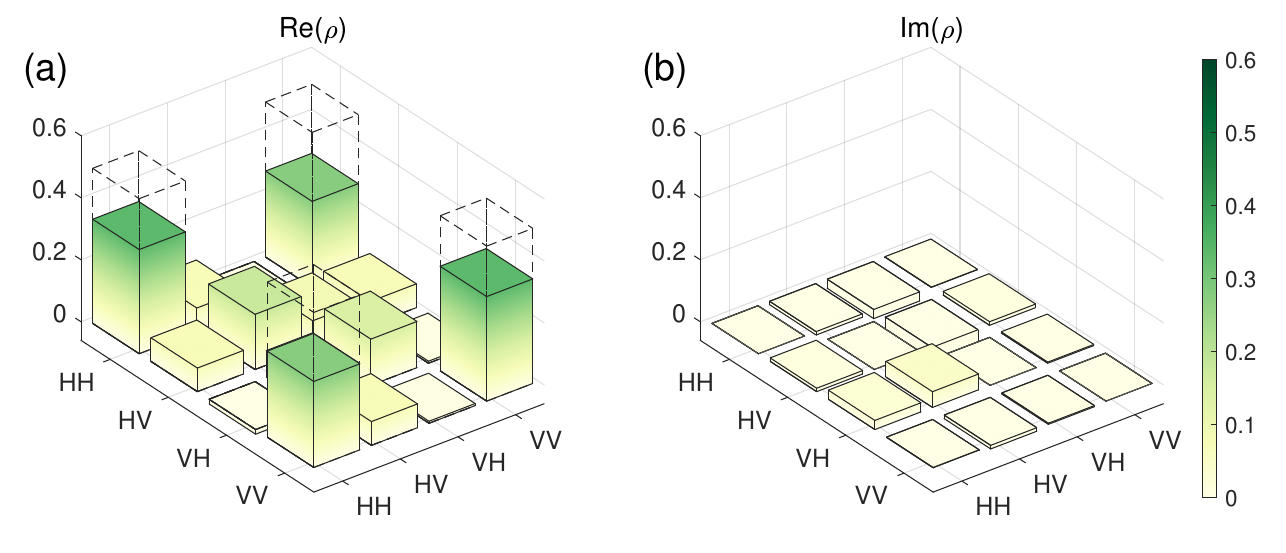}
\caption{Quantum-state-tomography results conditioned on the detection of photon 2 in $\ket{\rm V}_2$ only. (a) Real and (b) the imaginary part of the density matrix of state $\ket{\psi}_{14}$. The experimental quantum state has a fidelity of $0.614\pm0.011$ with respect to the Bell state $\ket{\phi^+}=\frac{1}{\sqrt{2}}(\ket{\rm HH}+\ket{\rm VV})$. The dashed wire grid represents the ideal case for state $\ket{\phi^+}$. Uncertainties are obtained from 100 Monte Carlo simulations with counting Poisson statistics.}
\end{figure}

The quantum-state-tomography result of quantum state of photons 1 and 4, $\ket{\psi}_{14}$, conditioned on the detection of photon 2 is shown in Fig. 5. We obtain a fidelity of $F=0.614\pm 0.011$ with respect to the Bell state $\ket{\phi^+}$. With the witness operator $\mathcal{W}=0.5\hat{\mathcal{I}}- \ket{\phi^+}\bra{\phi^+}$, we get $\rm{Tr}(\rm{\mathcal{W}}\rho_{AB})=-0.114 \pm 0.011<0$ and the concurrence of the experimentally obtained state is $0.265 \pm 0.020$, we prove the entanglement between Alice and Bob \cite{PhysRevLett.77.1413,PhysRevLett.92.087902}. This shows the existence of quantum entanglement between photons 1 and 4 conditionally only on the detection of photon 2. This result demonstrates that our method is conceptually distinct from entanglement swapping. Even in the presence of undetected ancillary photons, entanglement can still be generated through path identity.

\section{Conclusion}
In this work, we demonstrate the successful generation of entanglement between two photons with no direct interaction or any common past between them. In stark contrast to entanglement swapping, no prior entanglement generation or measurement is required in this process. Entanglement is generated using the indistinguishability of the origins of the photons. Another key advantage of our method compared to entanglement swapping is that we do not need to detect all the ancillary photons. By carefully controlling the intensity of pumps, we can establish entanglement even with one ancillary photon undetected. Our work paves the way for potential techniques in future quantum-communication and quantum-information applications.

From a fundamental perspective, entanglement swapping is based on BSM, which can be viewed as a quantum eraser. In such cases, a beam splitter and coincidence measurement of the two ancillary photons are used to remove distinguishable information. In contrast, our method relies on path identity, where distinguishable information is inherently absent, eliminating the need for any which-path information to be erased.

Our alternative process to generate entanglement between two never-interacting particles has applications in quantum communication in particular, specifically in reducing the resource requirements for implementing complex multi-node quantum networks. We have shown above, for instance, that our method requires only one ancillary photon, in contrast to two photons that are required for entanglement swapping. In \cite{RuizGonzalez2023digitaldiscoveryof} (Example 77), a generalization of this technique is demonstrated that can further reduce the number of required particles for multi-pair entanglement swapping. Therefore, our work hints at the possibility that other quantum networks or communication protocols might be implemented with alternative or lower resource requirements. Interesting future directions involve the analysis of resource requirements and reductions in complex quantum teleportation tasks \cite{Wang2015,PhysRevLett.123.070505,PhysRevLett.125.230501}, and quantum networks with complex structures \cite{PhysRevA.97.032312,Hu2020}.

Although the present work focuses on the novel principle and implementation of entanglement generation, we also think to devise new and practical quantum information protocols, based on the principle presented in this work, is important and will study that in the future work. Note that a related but distinct work \cite{PhysRevA.106.032609} has shown that, by using a linear optical network consisting of beam splitters and beam displacers, one entangles two independent photons.
\\
\\

We thank A. Zeilinger for fruitful discussions. This research was supported by the National Key Research and Development Program of China (Grants Nos. 2022YFE0137000, 2019YFA0308704), Natural Science Foundation of Jiangsu Province (Grants Nos. BK20240006, BK20233001), the Leading-Edge Technology Program of Jiangsu Natural Science Foundation (Grant No. BK20192001), the Fundamental Research Funds for the Central Universities (Grant No. 2024300324), the Innovation Program for Quantum Science and Technology (Grants Nos. 2021ZD0300700, 2021ZD0301500), the Jiangsu Funding Program for Excellent Postdoctoral Talent (No. 20220ZB60), and the National Natural Science Foundation of China (Grant no. 12304397).

\appendix
\renewcommand{\theequation}{A\arabic{equation}} 
\renewcommand{\thefigure}{A\arabic{figure}} 
\renewcommand{\thetable}{A\arabic{table}} 
\setcounter{equation}{0} 
\setcounter{figure}{0} 
\setcounter{table}{0} 

\section{A1. Theoretical derivations of entangling independent particles using quantum indistinguishability}
As each source generates polarization-orthogonal photons, the two-mode squeezing operator is $S(\varepsilon)=e^{\varepsilon(\hat{a}^{\dagger}_{\rm iH}\hat{a}^{\dagger}_{\rm jV}-\hat{a}_{\rm iH}\hat{a}_{\rm jV})}$, where $\varepsilon$ quantifies the two-mode squeezing strength and is related to the generation efficiency of down-converted photons. For the Spontaneous Parametric Down-Conversion (SPDC) process, $\varepsilon$ is a small value. Here we use the approximate form of S: $S(\varepsilon)=\hat{\rm I}+\varepsilon(\hat{a}^{\dagger}_{\rm iH}\hat{a}^{\dagger}_{\rm jV}-\hat{a}_{\rm iH}\hat{a}_{\rm jV})$ . We omit the higher orders of $\varepsilon$, as they represent the multi-pair generation process. 

With the two pumps $\rm PB_1$ and $\rm PB_2$ (see Fig. 1(b)), we generate the initial state:

\begin{align}
\nonumber \ket{\psi}=&[\hat{\mathrm{I}}+\varepsilon _1(\hat{a}_{1\mathrm{H}}^{\dagger}\hat{a}_{3\mathrm{V}}^{\dagger}-\hat{a}_{1\mathrm{H}}\hat{a}_{3\mathrm{V}})]_{\mathrm{P}_1}\\
&\cdot [\hat{\mathrm{I}}+\varepsilon _2(\hat{a}_{2\mathrm{V}}^{\dagger}\hat{a}_{4\mathrm{H}}^{\dagger}-\hat{a}_{2\mathrm{V}}\hat{a}_{4\mathrm{H}})]_{\mathrm{P}_2}\ket{\rm vac},
\end{align}
where $\varepsilon_1$ and $\varepsilon_2$  represent the generation efficiency of sources $\rm P_1$ ($\rm i_1$, $\rm s_1$) and $\rm P_2$ ($\rm i_2$, $\rm s_2$).

After sources $\rm P_1$ and $\rm P_2$, we insert two polarization rotators on the paths of 1 and 4, which implement the operations:

\begin{align}
\nonumber &\hat{a}_{1\mathrm{V}}^{\dagger}\rightarrow \hat{a}_{1\mathrm{H}}^{\dagger},\hat{a}_{4\mathrm{V}}^{\dagger}\rightarrow \hat{a}_{4\mathrm{H}}^{\dagger},\\
&\hat{a}_{1\mathrm{H}}^{\dagger}\rightarrow \hat{a}_{1\mathrm{V}}^{\dagger},\hat{a}_{4\mathrm{H}}^{\dagger}\rightarrow \hat{a}_{4\mathrm{V}}^{\dagger},
\end{align}
and change the initial state to

\begin{equation}
|\psi'\rangle =[\hat{\mathrm{I}}+\varepsilon _1\hat{a}_{1\mathrm{V}}^{\dagger}\hat{a}_{3\mathrm{V}}^{\dagger}]_{\mathrm{P}_1}\cdot [\hat{\mathrm{I}}+\varepsilon _2\hat{a}_{2\mathrm{V}}^{\dagger}\hat{a}_{4\mathrm{V}}^{\dagger}]_{\mathrm{P}_2}|\mathrm{vac}\rangle.
\end{equation}

Then we add the other two sources $\rm P_3$ and $\rm P_4$, and obtain the final state:
\begin{align}
\nonumber |\psi \rangle _f&=[\hat{\mathrm{I}}+\varepsilon _3(\hat{a}_{1\mathrm{H}}^{\dagger}\hat{a}_{2\mathrm{V}}^{\dagger}-\hat{a}_{1\mathrm{H}}\hat{a}_{2\mathrm{V}})]_{\mathrm{P}_3}\\
\nonumber &\cdot [\hat{\mathrm{I}}+\varepsilon _4(\hat{a}_{3\mathrm{V}}^{\dagger}\hat{a}_{4\mathrm{H}}^{\dagger}-\hat{a}_{3\mathrm{V}}\hat{a}_{4\mathrm{H}})]_{\mathrm{P}_4}|\psi \prime\rangle 
\\
\nonumber &=\ket{\rm vac} +[\varepsilon _1\hat{a}_{1\mathrm{V}}^{\dagger}\hat{a}_{3\mathrm{V}}^{\dagger}+\varepsilon _2\hat{a}_{2\mathrm{V}}^{\dagger}\hat{a}_{4\mathrm{V}}^{\dagger}\\
\nonumber &+\varepsilon _3\hat{a}_{1\mathrm{H}}^{\dagger}\hat{a}_{2\mathrm{V}}^{\dagger}+\varepsilon _4\hat{a}_{3\mathrm{V}}^{\dagger}\hat{a}_{4\mathrm{H}}^{\dagger}]\ket{\rm vac} 
\\
\nonumber &+[\varepsilon _1\varepsilon _2\hat{a}_{1\mathrm{V}}^{\dagger}\hat{a}_{3\mathrm{V}}^{\dagger}\hat{a}_{2\mathrm{V}}^{\dagger}\hat{a}_{4\mathrm{V}}^{\dagger}+\varepsilon _3\varepsilon _4\hat{a}_{1\mathrm{H}}^{\dagger}\hat{a}_{2\mathrm{V}}^{\dagger}\hat{a}_{3\mathrm{V}}^{\dagger}\hat{a}_{4\mathrm{H}}^{\dagger}
\\
\nonumber &+\varepsilon _1\varepsilon _3\hat{a}_{1\mathrm{V}}^{\dagger}\hat{a}_{3\mathrm{V}}^{\dagger}\hat{a}_{1\mathrm{H}}^{\dagger}\hat{a}_{2\mathrm{V}}^{\dagger}+\varepsilon _1\varepsilon _4\hat{a}_{1\mathrm{V}}^{\dagger}\hat{a}_{3\mathrm{V}}^{\dagger}\hat{a}_{3\mathrm{V}}^{\dagger}\hat{a}_{4\mathrm{H}}^{\dagger}\\
&+\varepsilon _2\varepsilon _3\hat{a}_{2\mathrm{V}}^{\dagger}\hat{a}_{4\mathrm{V}}^{\dagger}\hat{a}_{1\mathrm{H}}^{\dagger}\hat{a}_{2\mathrm{V}}^{\dagger}+\varepsilon _2\varepsilon _4\hat{a}_{2\mathrm{V}}^{\dagger}\hat{a}_{4\mathrm{V}}^{\dagger}\hat{a}_{3\mathrm{V}}^{\dagger}\hat{a}_{4\mathrm{H}}^{\dagger}]\ket{\rm vac} 
\end{align}

The second-order terms of Eq. (A4) form Eq. (1) in the main text.

\section{A2. Complete data of joint probability distributions of photons 1 and 4 in different bases}

Fig. A1 shows the complete joint probability distributions of photons 1 and 4 in different bases.

\section{A3. Determining the relative phase of the entangled states between photons 1 and 4}

In Eq. 2 in the main text, as the pumps and down-conversion photons all experience their respective phases, there is a relative phase $\delta$ between the two superposed terms in the entangled state $\ket{\psi}=(\ket{\rm HH}+e^{i\delta}\ket{\rm VV})_{14}\ket{\rm VV}_{23}$. We tune the phase of the idler photons $\phi_i$ and observe the interference between Alice and Bob in the D/A bases, as shown in Fig. A2. When the counts of DD and AA are maximal, and the counts of DA and AD are minimal, we have the entangled state $\ket{\phi^+}=\frac{1}{\sqrt{2}}(\ket{\rm HH}+\ket{\rm VV})$ and $\delta=0$. Therefore, we can control the phase $\phi_i$ of the idler photons ($\rm i_1$, $\rm i_2$) to set $\delta$ to the zero point.

\section{A4. Determinations of photon-pair generation efficiencies of four SPDC sources}

 In the experiment, we adjust the generation efficiency of the SPDC by varying the pump powers ($\rm PB_1$–$\rm PB_4$), while keeping the couplers' positions unchanged. This ensures that the coupling efficiencies of the photons remain constant. We record the changes in the coincidence counting rate and monitor the variations in the generation efficiencies of the four sources, which enables us to calculate the ratio $\frac{\varepsilon'}{\varepsilon}$. We measure the coincidence counting rates of the four two-photon sources: Source I ($\rm s_1$, $\rm i_1$) at $\sim$32.9 kHz, source II ($\rm s_2$, $\rm i_2$) at $\sim$1.0 kHz, source III ($\rm s_3$, $\rm i_3$) at $\sim$1.2 kHz, and source IV ($\rm s_4$, $\rm i_4$) at $\sim$33.3 kHz. Therefore, we obtain:
 \begin{equation}
 \frac{\varepsilon'}{\varepsilon}\approx \sqrt{\frac{\varepsilon _2\varepsilon _3}{\varepsilon _1\varepsilon _4}}=(\frac{\mathrm{CC}_2\cdot \mathrm{CC}_3}{\mathrm{CC}_1\cdot \mathrm{CC}_4})^{\frac{1}{4}}=0.182
 \end{equation}

\begin{figure}
\includegraphics[width=8cm]{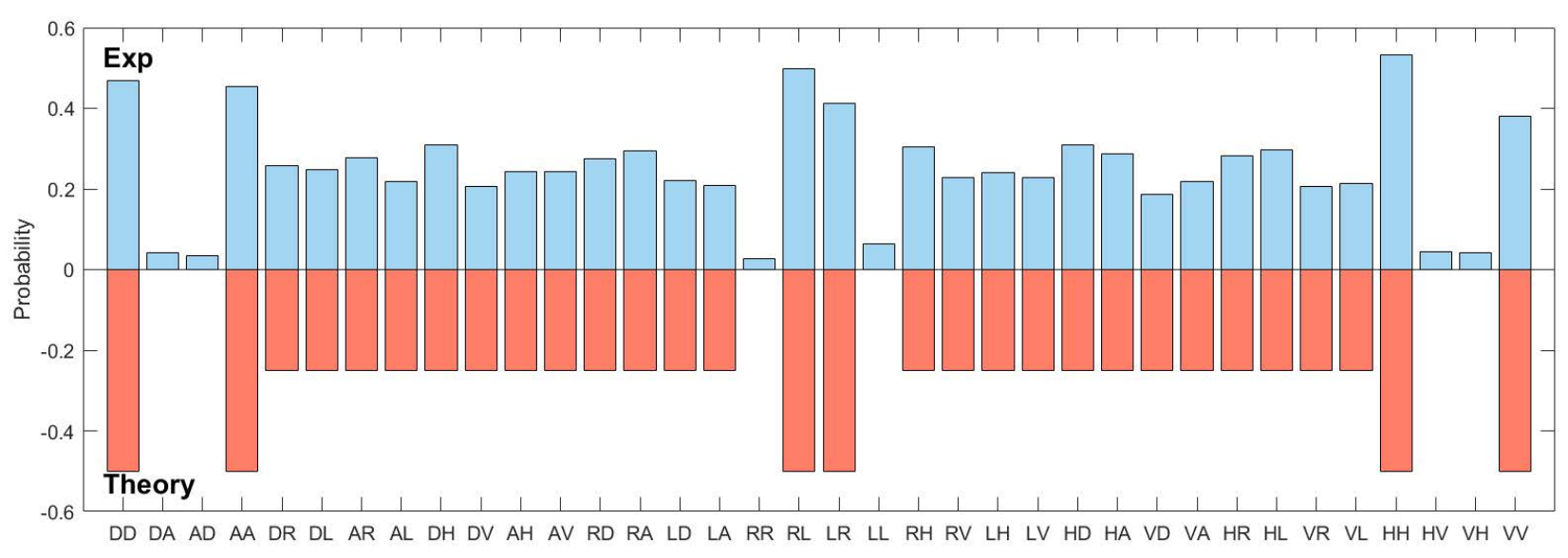}
\caption{Complete joint probability distributions of photons 1 and 4 in different bases. The experiment results (blue) compared with theoretical predictions (red). Alice and Bob measure their photons in H/V, D/A, R/L bases, and obtain their coincidence counts conditioned on the coincidence detection of photons 2 and 3.}
\end{figure}

\begin{figure}
\includegraphics[width=9cm]{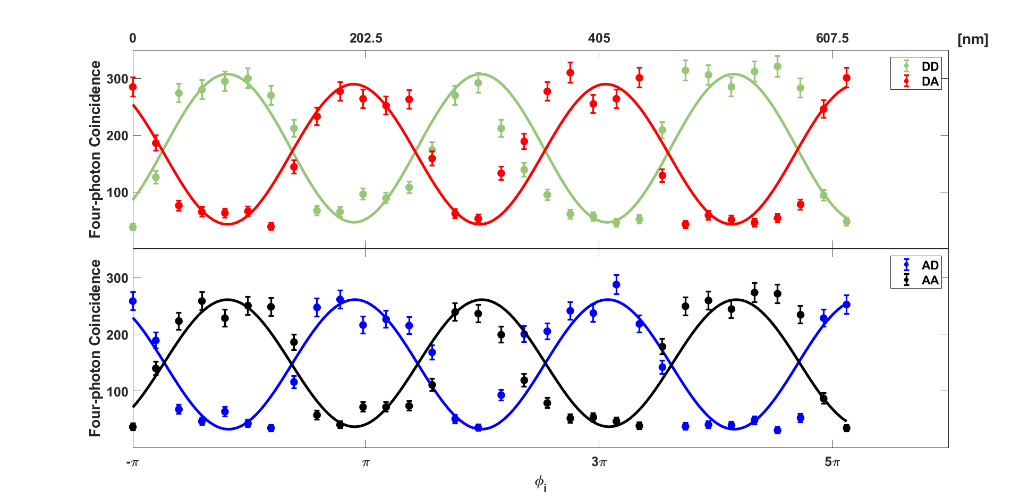}
\caption{The coincidence counts of Alice and Bob in the diagonal (D) and anti-diagonal polarization bases (A). The interference visibilities of the four curves are $0.7883\pm0.0325$ (DD), $0.7765\pm0.0337$ (DA), $0.8050\pm0.0333$ (AD), and $0.7727\pm0.0362$ (AA). Although the phase of this four-photon interferometer is passively stabilized with surrounding enclosure, the instability of the coincidence counts can still be seen in the deviations of experimental results from the theoretical fittings.}
\end{figure}

\clearpage

\begin{thebibliography}{60}%
\makeatletter
\providecommand \@ifxundefined [1]{%
 \@ifx{#1\undefined}
}%
\providecommand \@ifnum [1]{%
 \ifnum #1\expandafter \@firstoftwo
 \else \expandafter \@secondoftwo
 \fi
}%
\providecommand \@ifx [1]{%
 \ifx #1\expandafter \@firstoftwo
 \else \expandafter \@secondoftwo
 \fi
}%
\providecommand \natexlab [1]{#1}%
\providecommand \enquote  [1]{``#1''}%
\providecommand \bibnamefont  [1]{#1}%
\providecommand \bibfnamefont [1]{#1}%
\providecommand \citenamefont [1]{#1}%
\providecommand \href@noop [0]{\@secondoftwo}%
\providecommand \href [0]{\begingroup \@sanitize@url \@href}%
\providecommand \@href[1]{\@@startlink{#1}\@@href}%
\providecommand \@@href[1]{\endgroup#1\@@endlink}%
\providecommand \@sanitize@url [0]{\catcode `\\12\catcode `\$12\catcode
  `\&12\catcode `\#12\catcode `\^12\catcode `\_12\catcode `\%12\relax}%
\providecommand \@@startlink[1]{}%
\providecommand \@@endlink[0]{}%
\providecommand \url  [0]{\begingroup\@sanitize@url \@url }%
\providecommand \@url [1]{\endgroup\@href {#1}{\urlprefix }}%
\providecommand \urlprefix  [0]{URL }%
\providecommand \Eprint [0]{\href }%
\providecommand \doibase [0]{https://doi.org/}%
\providecommand \selectlanguage [0]{\@gobble}%
\providecommand \bibinfo  [0]{\@secondoftwo}%
\providecommand \bibfield  [0]{\@secondoftwo}%
\providecommand \translation [1]{[#1]}%
\providecommand \BibitemOpen [0]{}%
\providecommand \bibitemStop [0]{}%
\providecommand \bibitemNoStop [0]{.\EOS\space}%
\providecommand \EOS [0]{\spacefactor3000\relax}%
\providecommand \BibitemShut  [1]{\csname bibitem#1\endcsname}%
\let\auto@bib@innerbib\@empty
\bibitem [{\citenamefont {Einstein}\ \emph {et~al.}(1935)\citenamefont
  {Einstein}, \citenamefont {Podolsky},\ and\ \citenamefont
  {Rosen}}]{PhysRev.47.777}%
  \BibitemOpen
  \bibfield  {author} {\bibinfo {author} {\bibfnamefont {A.}~\bibnamefont
  {Einstein}}, \bibinfo {author} {\bibfnamefont {B.}~\bibnamefont {Podolsky}},\
  and\ \bibinfo {author} {\bibfnamefont {N.}~\bibnamefont {Rosen}},\ }\bibfield
   {title} {\bibinfo {title} {Can quantum-mechanical description of physical
  reality be considered complete?},\ }\href
  {https://doi.org/10.1103/PhysRev.47.777} {\bibfield  {journal} {\bibinfo
  {journal} {Phys. Rev.}\ }\textbf {\bibinfo {volume} {47}},\ \bibinfo {pages}
  {777} (\bibinfo {year} {1935})}\BibitemShut {NoStop}%
\bibitem [{\citenamefont {Bohr}(1935)}]{PhysRev.48.696}%
  \BibitemOpen
  \bibfield  {author} {\bibinfo {author} {\bibfnamefont {N.}~\bibnamefont
  {Bohr}},\ }\bibfield  {title} {\bibinfo {title} {Can quantum-mechanical
  description of physical reality be considered complete?},\ }\href
  {https://doi.org/10.1103/PhysRev.48.696} {\bibfield  {journal} {\bibinfo
  {journal} {Phys. Rev.}\ }\textbf {\bibinfo {volume} {48}},\ \bibinfo {pages}
  {696} (\bibinfo {year} {1935})}\BibitemShut {NoStop}%
\bibitem [{\citenamefont {Schr{\"o}dinger}(1935)}]{Schrödinger1935}%
  \BibitemOpen
  \bibfield  {author} {\bibinfo {author} {\bibfnamefont {E.}~\bibnamefont
  {Schr{\"o}dinger}},\ }\bibfield  {title} {\bibinfo {title} {Die
  gegenw{\"a}rtige situation in der quantenmechanik},\ }\href
  {https://doi.org/10.1007/BF01491987} {\bibfield  {journal} {\bibinfo
  {journal} {Naturwissenschaften}\ }\textbf {\bibinfo {volume} {23}},\ \bibinfo
  {pages} {844} (\bibinfo {year} {1935})}\BibitemShut {NoStop}%
\bibitem [{\citenamefont {Wu}\ and\ \citenamefont
  {Shaknov}(1950)}]{PhysRev.77.136}%
  \BibitemOpen
  \bibfield  {author} {\bibinfo {author} {\bibfnamefont {C.~S.}\ \bibnamefont
  {Wu}}\ and\ \bibinfo {author} {\bibfnamefont {I.}~\bibnamefont {Shaknov}},\
  }\bibfield  {title} {\bibinfo {title} {The angular correlation of scattered
  annihilation radiation},\ }\href {https://doi.org/10.1103/PhysRev.77.136}
  {\bibfield  {journal} {\bibinfo  {journal} {Phys. Rev.}\ }\textbf {\bibinfo
  {volume} {77}},\ \bibinfo {pages} {136} (\bibinfo {year} {1950})}\BibitemShut
  {NoStop}%
\bibitem [{\citenamefont {Freedman}\ and\ \citenamefont
  {Clauser}(1972)}]{PhysRevLett.28.938}%
  \BibitemOpen
  \bibfield  {author} {\bibinfo {author} {\bibfnamefont {S.~J.}\ \bibnamefont
  {Freedman}}\ and\ \bibinfo {author} {\bibfnamefont {J.~F.}\ \bibnamefont
  {Clauser}},\ }\bibfield  {title} {\bibinfo {title} {Experimental test of
  local hidden-variable theories},\ }\href
  {https://doi.org/10.1103/PhysRevLett.28.938} {\bibfield  {journal} {\bibinfo
  {journal} {Phys. Rev. Lett.}\ }\textbf {\bibinfo {volume} {28}},\ \bibinfo
  {pages} {938} (\bibinfo {year} {1972})}\BibitemShut {NoStop}%
\bibitem [{\citenamefont {Aspect}\ \emph {et~al.}(1982)\citenamefont {Aspect},
  \citenamefont {Dalibard},\ and\ \citenamefont {Roger}}]{PhysRevLett.49.1804}%
  \BibitemOpen
  \bibfield  {author} {\bibinfo {author} {\bibfnamefont {A.}~\bibnamefont
  {Aspect}}, \bibinfo {author} {\bibfnamefont {J.}~\bibnamefont {Dalibard}},\
  and\ \bibinfo {author} {\bibfnamefont {G.}~\bibnamefont {Roger}},\ }\bibfield
   {title} {\bibinfo {title} {Experimental test of bell's inequalities using
  time-varying analyzers},\ }\href
  {https://doi.org/10.1103/PhysRevLett.49.1804} {\bibfield  {journal} {\bibinfo
   {journal} {Phys. Rev. Lett.}\ }\textbf {\bibinfo {volume} {49}},\ \bibinfo
  {pages} {1804} (\bibinfo {year} {1982})}\BibitemShut {NoStop}%
\bibitem [{\citenamefont {Shih}\ and\ \citenamefont
  {Alley}(1988)}]{PhysRevLett.61.2921}%
  \BibitemOpen
  \bibfield  {author} {\bibinfo {author} {\bibfnamefont {Y.~H.}\ \bibnamefont
  {Shih}}\ and\ \bibinfo {author} {\bibfnamefont {C.~O.}\ \bibnamefont
  {Alley}},\ }\bibfield  {title} {\bibinfo {title} {New type of
  einstein-podolsky-rosen-bohm experiment using pairs of light quanta produced
  by optical parametric down conversion},\ }\href
  {https://doi.org/10.1103/PhysRevLett.61.2921} {\bibfield  {journal} {\bibinfo
   {journal} {Phys. Rev. Lett.}\ }\textbf {\bibinfo {volume} {61}},\ \bibinfo
  {pages} {2921} (\bibinfo {year} {1988})}\BibitemShut {NoStop}%
\bibitem [{\citenamefont {Ou}\ and\ \citenamefont
  {Mandel}(1988)}]{PhysRevLett.61.50}%
  \BibitemOpen
  \bibfield  {author} {\bibinfo {author} {\bibfnamefont {Z.~Y.}\ \bibnamefont
  {Ou}}\ and\ \bibinfo {author} {\bibfnamefont {L.}~\bibnamefont {Mandel}},\
  }\bibfield  {title} {\bibinfo {title} {Violation of bell's inequality and
  classical probability in a two-photon correlation experiment},\ }\href
  {https://doi.org/10.1103/PhysRevLett.61.50} {\bibfield  {journal} {\bibinfo
  {journal} {Phys. Rev. Lett.}\ }\textbf {\bibinfo {volume} {61}},\ \bibinfo
  {pages} {50} (\bibinfo {year} {1988})}\BibitemShut {NoStop}%
\bibitem [{\citenamefont {Kiess}\ \emph {et~al.}(1993)\citenamefont {Kiess},
  \citenamefont {Shih}, \citenamefont {Sergienko},\ and\ \citenamefont
  {Alley}}]{PhysRevLett.71.3893}%
  \BibitemOpen
  \bibfield  {author} {\bibinfo {author} {\bibfnamefont {T.~E.}\ \bibnamefont
  {Kiess}}, \bibinfo {author} {\bibfnamefont {Y.~H.}\ \bibnamefont {Shih}},
  \bibinfo {author} {\bibfnamefont {A.~V.}\ \bibnamefont {Sergienko}},\ and\
  \bibinfo {author} {\bibfnamefont {C.~O.}\ \bibnamefont {Alley}},\ }\bibfield
  {title} {\bibinfo {title} {Einstein-podolsky-rosen-bohm experiment using
  pairs of light quanta produced by type-ii parametric down-conversion},\
  }\href {https://doi.org/10.1103/PhysRevLett.71.3893} {\bibfield  {journal}
  {\bibinfo  {journal} {Phys. Rev. Lett.}\ }\textbf {\bibinfo {volume} {71}},\
  \bibinfo {pages} {3893} (\bibinfo {year} {1993})}\BibitemShut {NoStop}%
\bibitem [{\citenamefont {Hardy}(1992)}]{HARDY1992326}%
  \BibitemOpen
  \bibfield  {author} {\bibinfo {author} {\bibfnamefont {L.}~\bibnamefont
  {Hardy}},\ }\bibfield  {title} {\bibinfo {title} {Source of photons with
  correlated polarisations and correlated directions},\ }\href
  {https://doi.org/https://doi.org/10.1016/0375-9601(92)90554-Y} {\bibfield
  {journal} {\bibinfo  {journal} {Phys. Lett. A}\ }\textbf {\bibinfo {volume}
  {161}},\ \bibinfo {pages} {326} (\bibinfo {year} {1992})}\BibitemShut
  {NoStop}%
\bibitem [{\citenamefont {Kwiat}\ \emph {et~al.}(1995)\citenamefont {Kwiat},
  \citenamefont {Mattle}, \citenamefont {Weinfurter}, \citenamefont
  {Zeilinger}, \citenamefont {Sergienko},\ and\ \citenamefont
  {Shih}}]{PhysRevLett.75.4337}%
  \BibitemOpen
  \bibfield  {author} {\bibinfo {author} {\bibfnamefont {P.~G.}\ \bibnamefont
  {Kwiat}}, \bibinfo {author} {\bibfnamefont {K.}~\bibnamefont {Mattle}},
  \bibinfo {author} {\bibfnamefont {H.}~\bibnamefont {Weinfurter}}, \bibinfo
  {author} {\bibfnamefont {A.}~\bibnamefont {Zeilinger}}, \bibinfo {author}
  {\bibfnamefont {A.~V.}\ \bibnamefont {Sergienko}},\ and\ \bibinfo {author}
  {\bibfnamefont {Y.}~\bibnamefont {Shih}},\ }\bibfield  {title} {\bibinfo
  {title} {New high-intensity source of polarization-entangled photon pairs},\
  }\href {https://doi.org/10.1103/PhysRevLett.75.4337} {\bibfield  {journal}
  {\bibinfo  {journal} {Phys. Rev. Lett.}\ }\textbf {\bibinfo {volume} {75}},\
  \bibinfo {pages} {4337} (\bibinfo {year} {1995})}\BibitemShut {NoStop}%
\bibitem [{\citenamefont {Giustina}\ \emph {et~al.}(2015)\citenamefont
  {Giustina}, \citenamefont {Versteegh}, \citenamefont {Wengerowsky},
  \citenamefont {Handsteiner}, \citenamefont {Hochrainer}, \citenamefont
  {Phelan}, \citenamefont {Steinlechner}, \citenamefont {Kofler}, \citenamefont
  {Larsson}, \citenamefont {Abell\'an}, \citenamefont {Amaya}, \citenamefont
  {Pruneri}, \citenamefont {Mitchell}, \citenamefont {Beyer}, \citenamefont
  {Gerrits}, \citenamefont {Lita}, \citenamefont {Shalm}, \citenamefont {Nam},
  \citenamefont {Scheidl}, \citenamefont {Ursin}, \citenamefont {Wittmann},\
  and\ \citenamefont {Zeilinger}}]{PhysRevLett.115.250401}%
  \BibitemOpen
  \bibfield  {author} {\bibinfo {author} {\bibfnamefont {M.}~\bibnamefont
  {Giustina}}, \bibinfo {author} {\bibfnamefont {M.~A.~M.}\ \bibnamefont
  {Versteegh}}, \bibinfo {author} {\bibfnamefont {S.}~\bibnamefont
  {Wengerowsky}}, \bibinfo {author} {\bibfnamefont {J.}~\bibnamefont
  {Handsteiner}}, \bibinfo {author} {\bibfnamefont {A.}~\bibnamefont
  {Hochrainer}}, \bibinfo {author} {\bibfnamefont {K.}~\bibnamefont {Phelan}},
  \bibinfo {author} {\bibfnamefont {F.}~\bibnamefont {Steinlechner}}, \bibinfo
  {author} {\bibfnamefont {J.}~\bibnamefont {Kofler}}, \bibinfo {author}
  {\bibfnamefont {J.-A.}\ \bibnamefont {Larsson}}, \bibinfo {author}
  {\bibfnamefont {C.}~\bibnamefont {Abell\'an}}, \bibinfo {author}
  {\bibfnamefont {W.}~\bibnamefont {Amaya}}, \bibinfo {author} {\bibfnamefont
  {V.}~\bibnamefont {Pruneri}}, \bibinfo {author} {\bibfnamefont {M.~W.}\
  \bibnamefont {Mitchell}}, \bibinfo {author} {\bibfnamefont {J.}~\bibnamefont
  {Beyer}}, \bibinfo {author} {\bibfnamefont {T.}~\bibnamefont {Gerrits}},
  \bibinfo {author} {\bibfnamefont {A.~E.}\ \bibnamefont {Lita}}, \bibinfo
  {author} {\bibfnamefont {L.~K.}\ \bibnamefont {Shalm}}, \bibinfo {author}
  {\bibfnamefont {S.~W.}\ \bibnamefont {Nam}}, \bibinfo {author} {\bibfnamefont
  {T.}~\bibnamefont {Scheidl}}, \bibinfo {author} {\bibfnamefont
  {R.}~\bibnamefont {Ursin}}, \bibinfo {author} {\bibfnamefont
  {B.}~\bibnamefont {Wittmann}},\ and\ \bibinfo {author} {\bibfnamefont
  {A.}~\bibnamefont {Zeilinger}},\ }\bibfield  {title} {\bibinfo {title}
  {Significant-loophole-free test of bell's theorem with entangled photons},\
  }\href {https://doi.org/10.1103/PhysRevLett.115.250401} {\bibfield  {journal}
  {\bibinfo  {journal} {Phys. Rev. Lett.}\ }\textbf {\bibinfo {volume} {115}},\
  \bibinfo {pages} {250401} (\bibinfo {year} {2015})}\BibitemShut {NoStop}%
\bibitem [{\citenamefont {Shalm}\ \emph {et~al.}(2015)\citenamefont {Shalm},
  \citenamefont {Meyer-Scott}, \citenamefont {Christensen}, \citenamefont
  {Bierhorst}, \citenamefont {Wayne}, \citenamefont {Stevens}, \citenamefont
  {Gerrits}, \citenamefont {Glancy}, \citenamefont {Hamel}, \citenamefont
  {Allman}, \citenamefont {Coakley}, \citenamefont {Dyer}, \citenamefont
  {Hodge}, \citenamefont {Lita}, \citenamefont {Verma}, \citenamefont
  {Lambrocco}, \citenamefont {Tortorici}, \citenamefont {Migdall},
  \citenamefont {Zhang}, \citenamefont {Kumor}, \citenamefont {Farr},
  \citenamefont {Marsili}, \citenamefont {Shaw}, \citenamefont {Stern},
  \citenamefont {Abell\'an}, \citenamefont {Amaya}, \citenamefont {Pruneri},
  \citenamefont {Jennewein}, \citenamefont {Mitchell}, \citenamefont {Kwiat},
  \citenamefont {Bienfang}, \citenamefont {Mirin}, \citenamefont {Knill},\ and\
  \citenamefont {Nam}}]{PhysRevLett.115.250402}%
  \BibitemOpen
  \bibfield  {author} {\bibinfo {author} {\bibfnamefont {L.~K.}\ \bibnamefont
  {Shalm}}, \bibinfo {author} {\bibfnamefont {E.}~\bibnamefont {Meyer-Scott}},
  \bibinfo {author} {\bibfnamefont {B.~G.}\ \bibnamefont {Christensen}},
  \bibinfo {author} {\bibfnamefont {P.}~\bibnamefont {Bierhorst}}, \bibinfo
  {author} {\bibfnamefont {M.~A.}\ \bibnamefont {Wayne}}, \bibinfo {author}
  {\bibfnamefont {M.~J.}\ \bibnamefont {Stevens}}, \bibinfo {author}
  {\bibfnamefont {T.}~\bibnamefont {Gerrits}}, \bibinfo {author} {\bibfnamefont
  {S.}~\bibnamefont {Glancy}}, \bibinfo {author} {\bibfnamefont {D.~R.}\
  \bibnamefont {Hamel}}, \bibinfo {author} {\bibfnamefont {M.~S.}\ \bibnamefont
  {Allman}}, \bibinfo {author} {\bibfnamefont {K.~J.}\ \bibnamefont {Coakley}},
  \bibinfo {author} {\bibfnamefont {S.~D.}\ \bibnamefont {Dyer}}, \bibinfo
  {author} {\bibfnamefont {C.}~\bibnamefont {Hodge}}, \bibinfo {author}
  {\bibfnamefont {A.~E.}\ \bibnamefont {Lita}}, \bibinfo {author}
  {\bibfnamefont {V.~B.}\ \bibnamefont {Verma}}, \bibinfo {author}
  {\bibfnamefont {C.}~\bibnamefont {Lambrocco}}, \bibinfo {author}
  {\bibfnamefont {E.}~\bibnamefont {Tortorici}}, \bibinfo {author}
  {\bibfnamefont {A.~L.}\ \bibnamefont {Migdall}}, \bibinfo {author}
  {\bibfnamefont {Y.}~\bibnamefont {Zhang}}, \bibinfo {author} {\bibfnamefont
  {D.~R.}\ \bibnamefont {Kumor}}, \bibinfo {author} {\bibfnamefont {W.~H.}\
  \bibnamefont {Farr}}, \bibinfo {author} {\bibfnamefont {F.}~\bibnamefont
  {Marsili}}, \bibinfo {author} {\bibfnamefont {M.~D.}\ \bibnamefont {Shaw}},
  \bibinfo {author} {\bibfnamefont {J.~A.}\ \bibnamefont {Stern}}, \bibinfo
  {author} {\bibfnamefont {C.}~\bibnamefont {Abell\'an}}, \bibinfo {author}
  {\bibfnamefont {W.}~\bibnamefont {Amaya}}, \bibinfo {author} {\bibfnamefont
  {V.}~\bibnamefont {Pruneri}}, \bibinfo {author} {\bibfnamefont
  {T.}~\bibnamefont {Jennewein}}, \bibinfo {author} {\bibfnamefont {M.~W.}\
  \bibnamefont {Mitchell}}, \bibinfo {author} {\bibfnamefont {P.~G.}\
  \bibnamefont {Kwiat}}, \bibinfo {author} {\bibfnamefont {J.~C.}\ \bibnamefont
  {Bienfang}}, \bibinfo {author} {\bibfnamefont {R.~P.}\ \bibnamefont {Mirin}},
  \bibinfo {author} {\bibfnamefont {E.}~\bibnamefont {Knill}},\ and\ \bibinfo
  {author} {\bibfnamefont {S.~W.}\ \bibnamefont {Nam}},\ }\bibfield  {title}
  {\bibinfo {title} {Strong loophole-free test of local realism},\ }\href
  {https://doi.org/10.1103/PhysRevLett.115.250402} {\bibfield  {journal}
  {\bibinfo  {journal} {Phys. Rev. Lett.}\ }\textbf {\bibinfo {volume} {115}},\
  \bibinfo {pages} {250402} (\bibinfo {year} {2015})}\BibitemShut {NoStop}%
\bibitem [{\citenamefont {Hagley}\ \emph {et~al.}(1997)\citenamefont {Hagley},
  \citenamefont {Ma\^{\i}tre}, \citenamefont {Nogues}, \citenamefont
  {Wunderlich}, \citenamefont {Brune}, \citenamefont {Raimond},\ and\
  \citenamefont {Haroche}}]{PhysRevLett.79.1}%
  \BibitemOpen
  \bibfield  {author} {\bibinfo {author} {\bibfnamefont {E.}~\bibnamefont
  {Hagley}}, \bibinfo {author} {\bibfnamefont {X.}~\bibnamefont {Ma\^{\i}tre}},
  \bibinfo {author} {\bibfnamefont {G.}~\bibnamefont {Nogues}}, \bibinfo
  {author} {\bibfnamefont {C.}~\bibnamefont {Wunderlich}}, \bibinfo {author}
  {\bibfnamefont {M.}~\bibnamefont {Brune}}, \bibinfo {author} {\bibfnamefont
  {J.~M.}\ \bibnamefont {Raimond}},\ and\ \bibinfo {author} {\bibfnamefont
  {S.}~\bibnamefont {Haroche}},\ }\bibfield  {title} {\bibinfo {title}
  {Generation of einstein-podolsky-rosen pairs of atoms},\ }\href
  {https://doi.org/10.1103/PhysRevLett.79.1} {\bibfield  {journal} {\bibinfo
  {journal} {Phys. Rev. Lett.}\ }\textbf {\bibinfo {volume} {79}},\ \bibinfo
  {pages} {1} (\bibinfo {year} {1997})}\BibitemShut {NoStop}%
\bibitem [{\citenamefont {Julsgaard}\ \emph {et~al.}(2001)\citenamefont
  {Julsgaard}, \citenamefont {Kozhekin},\ and\ \citenamefont
  {Polzik}}]{Julsgaard2001}%
  \BibitemOpen
  \bibfield  {author} {\bibinfo {author} {\bibfnamefont {B.}~\bibnamefont
  {Julsgaard}}, \bibinfo {author} {\bibfnamefont {A.}~\bibnamefont
  {Kozhekin}},\ and\ \bibinfo {author} {\bibfnamefont {E.~S.}\ \bibnamefont
  {Polzik}},\ }\bibfield  {title} {\bibinfo {title} {Experimental long-lived
  entanglement of two macroscopic objects},\ }\href
  {https://doi.org/10.1038/35096524} {\bibfield  {journal} {\bibinfo  {journal}
  {Nature}\ }\textbf {\bibinfo {volume} {413}},\ \bibinfo {pages} {400}
  (\bibinfo {year} {2001})}\BibitemShut {NoStop}%
\bibitem [{\citenamefont {Majer}\ \emph {et~al.}(2007)\citenamefont {Majer},
  \citenamefont {Chow}, \citenamefont {Gambetta}, \citenamefont {Koch},
  \citenamefont {Johnson}, \citenamefont {Schreier}, \citenamefont {Frunzio},
  \citenamefont {Schuster}, \citenamefont {Houck}, \citenamefont {Wallraff},
  \citenamefont {Blais}, \citenamefont {Devoret}, \citenamefont {Girvin},\ and\
  \citenamefont {Schoelkopf}}]{Majer2007}%
  \BibitemOpen
  \bibfield  {author} {\bibinfo {author} {\bibfnamefont {J.}~\bibnamefont
  {Majer}}, \bibinfo {author} {\bibfnamefont {J.~M.}\ \bibnamefont {Chow}},
  \bibinfo {author} {\bibfnamefont {J.~M.}\ \bibnamefont {Gambetta}}, \bibinfo
  {author} {\bibfnamefont {J.}~\bibnamefont {Koch}}, \bibinfo {author}
  {\bibfnamefont {B.~R.}\ \bibnamefont {Johnson}}, \bibinfo {author}
  {\bibfnamefont {J.~A.}\ \bibnamefont {Schreier}}, \bibinfo {author}
  {\bibfnamefont {L.}~\bibnamefont {Frunzio}}, \bibinfo {author} {\bibfnamefont
  {D.~I.}\ \bibnamefont {Schuster}}, \bibinfo {author} {\bibfnamefont {A.~A.}\
  \bibnamefont {Houck}}, \bibinfo {author} {\bibfnamefont {A.}~\bibnamefont
  {Wallraff}}, \bibinfo {author} {\bibfnamefont {A.}~\bibnamefont {Blais}},
  \bibinfo {author} {\bibfnamefont {M.~H.}\ \bibnamefont {Devoret}}, \bibinfo
  {author} {\bibfnamefont {S.~M.}\ \bibnamefont {Girvin}},\ and\ \bibinfo
  {author} {\bibfnamefont {R.~J.}\ \bibnamefont {Schoelkopf}},\ }\bibfield
  {title} {\bibinfo {title} {Coupling superconducting qubits via a cavity
  bus},\ }\href {https://doi.org/10.1038/nature06184} {\bibfield  {journal}
  {\bibinfo  {journal} {Nature}\ }\textbf {\bibinfo {volume} {449}},\ \bibinfo
  {pages} {443} (\bibinfo {year} {2007})}\BibitemShut {NoStop}%
\bibitem [{\citenamefont {Jost}\ \emph {et~al.}(2009)\citenamefont {Jost},
  \citenamefont {Home}, \citenamefont {Amini}, \citenamefont {Hanneke},
  \citenamefont {Ozeri}, \citenamefont {Langer}, \citenamefont {Bollinger},
  \citenamefont {Leibfried},\ and\ \citenamefont {Wineland}}]{Jost2009}%
  \BibitemOpen
  \bibfield  {author} {\bibinfo {author} {\bibfnamefont {J.~D.}\ \bibnamefont
  {Jost}}, \bibinfo {author} {\bibfnamefont {J.~P.}\ \bibnamefont {Home}},
  \bibinfo {author} {\bibfnamefont {J.~M.}\ \bibnamefont {Amini}}, \bibinfo
  {author} {\bibfnamefont {D.}~\bibnamefont {Hanneke}}, \bibinfo {author}
  {\bibfnamefont {R.}~\bibnamefont {Ozeri}}, \bibinfo {author} {\bibfnamefont
  {C.}~\bibnamefont {Langer}}, \bibinfo {author} {\bibfnamefont {J.~J.}\
  \bibnamefont {Bollinger}}, \bibinfo {author} {\bibfnamefont {D.}~\bibnamefont
  {Leibfried}},\ and\ \bibinfo {author} {\bibfnamefont {D.~J.}\ \bibnamefont
  {Wineland}},\ }\bibfield  {title} {\bibinfo {title} {Entangled mechanical
  oscillators},\ }\href {https://doi.org/10.1038/nature08006} {\bibfield
  {journal} {\bibinfo  {journal} {Nature}\ }\textbf {\bibinfo {volume} {459}},\
  \bibinfo {pages} {683} (\bibinfo {year} {2009})}\BibitemShut {NoStop}%
\bibitem [{\citenamefont {Riedinger}\ \emph {et~al.}(2016)\citenamefont
  {Riedinger}, \citenamefont {Hong}, \citenamefont {Norte}, \citenamefont
  {Slater}, \citenamefont {Shang}, \citenamefont {Krause}, \citenamefont
  {Anant}, \citenamefont {Aspelmeyer},\ and\ \citenamefont
  {Gr{\"o}blacher}}]{Riedinger2016}%
  \BibitemOpen
  \bibfield  {author} {\bibinfo {author} {\bibfnamefont {R.}~\bibnamefont
  {Riedinger}}, \bibinfo {author} {\bibfnamefont {S.}~\bibnamefont {Hong}},
  \bibinfo {author} {\bibfnamefont {R.~A.}\ \bibnamefont {Norte}}, \bibinfo
  {author} {\bibfnamefont {J.~A.}\ \bibnamefont {Slater}}, \bibinfo {author}
  {\bibfnamefont {J.}~\bibnamefont {Shang}}, \bibinfo {author} {\bibfnamefont
  {A.~G.}\ \bibnamefont {Krause}}, \bibinfo {author} {\bibfnamefont
  {V.}~\bibnamefont {Anant}}, \bibinfo {author} {\bibfnamefont
  {M.}~\bibnamefont {Aspelmeyer}},\ and\ \bibinfo {author} {\bibfnamefont
  {S.}~\bibnamefont {Gr{\"o}blacher}},\ }\bibfield  {title} {\bibinfo {title}
  {Non-classical correlations between single photons and phonons from a
  mechanical oscillator},\ }\href {https://doi.org/10.1038/nature16536}
  {\bibfield  {journal} {\bibinfo  {journal} {Nature}\ }\textbf {\bibinfo
  {volume} {530}},\ \bibinfo {pages} {313} (\bibinfo {year}
  {2016})}\BibitemShut {NoStop}%
\bibitem [{\citenamefont {Ockeloen-Korppi}\ \emph {et~al.}(2018)\citenamefont
  {Ockeloen-Korppi}, \citenamefont {Damsk{\"a}gg}, \citenamefont
  {Pirkkalainen}, \citenamefont {Asjad}, \citenamefont {Clerk}, \citenamefont
  {Massel}, \citenamefont {Woolley},\ and\ \citenamefont
  {Sillanp{\"a}{\"a}}}]{Ockeloen-Korppi2018}%
  \BibitemOpen
  \bibfield  {author} {\bibinfo {author} {\bibfnamefont {C.~F.}\ \bibnamefont
  {Ockeloen-Korppi}}, \bibinfo {author} {\bibfnamefont {E.}~\bibnamefont
  {Damsk{\"a}gg}}, \bibinfo {author} {\bibfnamefont {J.-M.}\ \bibnamefont
  {Pirkkalainen}}, \bibinfo {author} {\bibfnamefont {M.}~\bibnamefont {Asjad}},
  \bibinfo {author} {\bibfnamefont {A.~A.}\ \bibnamefont {Clerk}}, \bibinfo
  {author} {\bibfnamefont {F.}~\bibnamefont {Massel}}, \bibinfo {author}
  {\bibfnamefont {M.~J.}\ \bibnamefont {Woolley}},\ and\ \bibinfo {author}
  {\bibfnamefont {M.~A.}\ \bibnamefont {Sillanp{\"a}{\"a}}},\ }\bibfield
  {title} {\bibinfo {title} {Stabilized entanglement of massive mechanical
  oscillators},\ }\href {https://doi.org/10.1038/s41586-018-0038-x} {\bibfield
  {journal} {\bibinfo  {journal} {Nature}\ }\textbf {\bibinfo {volume} {556}},\
  \bibinfo {pages} {478} (\bibinfo {year} {2018})}\BibitemShut {NoStop}%
\bibitem [{\citenamefont {Kurpiers}\ \emph {et~al.}(2018)\citenamefont
  {Kurpiers}, \citenamefont {Magnard}, \citenamefont {Walter}, \citenamefont
  {Royer}, \citenamefont {Pechal}, \citenamefont {Heinsoo}, \citenamefont
  {Salath{\'e}}, \citenamefont {Akin}, \citenamefont {Storz}, \citenamefont
  {Besse}, \citenamefont {Gasparinetti}, \citenamefont {Blais},\ and\
  \citenamefont {Wallraff}}]{Kurpiers2018}%
  \BibitemOpen
  \bibfield  {author} {\bibinfo {author} {\bibfnamefont {P.}~\bibnamefont
  {Kurpiers}}, \bibinfo {author} {\bibfnamefont {P.}~\bibnamefont {Magnard}},
  \bibinfo {author} {\bibfnamefont {T.}~\bibnamefont {Walter}}, \bibinfo
  {author} {\bibfnamefont {B.}~\bibnamefont {Royer}}, \bibinfo {author}
  {\bibfnamefont {M.}~\bibnamefont {Pechal}}, \bibinfo {author} {\bibfnamefont
  {J.}~\bibnamefont {Heinsoo}}, \bibinfo {author} {\bibfnamefont
  {Y.}~\bibnamefont {Salath{\'e}}}, \bibinfo {author} {\bibfnamefont
  {A.}~\bibnamefont {Akin}}, \bibinfo {author} {\bibfnamefont {S.}~\bibnamefont
  {Storz}}, \bibinfo {author} {\bibfnamefont {J.-C.}\ \bibnamefont {Besse}},
  \bibinfo {author} {\bibfnamefont {S.}~\bibnamefont {Gasparinetti}}, \bibinfo
  {author} {\bibfnamefont {A.}~\bibnamefont {Blais}},\ and\ \bibinfo {author}
  {\bibfnamefont {A.}~\bibnamefont {Wallraff}},\ }\bibfield  {title} {\bibinfo
  {title} {Deterministic quantum state transfer and remote entanglement using
  microwave photons},\ }\href {https://doi.org/10.1038/s41586-018-0195-y}
  {\bibfield  {journal} {\bibinfo  {journal} {Nature}\ }\textbf {\bibinfo
  {volume} {558}},\ \bibinfo {pages} {264} (\bibinfo {year}
  {2018})}\BibitemShut {NoStop}%
\bibitem [{\citenamefont {Bienfait}\ \emph {et~al.}(2019)\citenamefont
  {Bienfait}, \citenamefont {Satzinger}, \citenamefont {Zhong}, \citenamefont
  {Chang}, \citenamefont {Chou}, \citenamefont {Conner}, \citenamefont {Dumur},
  \citenamefont {Grebel}, \citenamefont {Peairs}, \citenamefont {Povey},\ and\
  \citenamefont {Cleland}}]{Bienfait2019}%
  \BibitemOpen
  \bibfield  {author} {\bibinfo {author} {\bibfnamefont {A.}~\bibnamefont
  {Bienfait}}, \bibinfo {author} {\bibfnamefont {K.~J.}\ \bibnamefont
  {Satzinger}}, \bibinfo {author} {\bibfnamefont {Y.~P.}\ \bibnamefont
  {Zhong}}, \bibinfo {author} {\bibfnamefont {H.-S.}\ \bibnamefont {Chang}},
  \bibinfo {author} {\bibfnamefont {M.-H.}\ \bibnamefont {Chou}}, \bibinfo
  {author} {\bibfnamefont {C.~R.}\ \bibnamefont {Conner}}, \bibinfo {author}
  {\bibfnamefont {{\'E}.}~\bibnamefont {Dumur}}, \bibinfo {author}
  {\bibfnamefont {J.}~\bibnamefont {Grebel}}, \bibinfo {author} {\bibfnamefont
  {G.~A.}\ \bibnamefont {Peairs}}, \bibinfo {author} {\bibfnamefont {R.~G.}\
  \bibnamefont {Povey}},\ and\ \bibinfo {author} {\bibfnamefont {A.~N.}\
  \bibnamefont {Cleland}},\ }\bibfield  {title} {\bibinfo {title}
  {Phonon-mediated quantum state transfer and remote qubit entanglement},\
  }\href {https://doi.org/10.1126/science.aaw8415} {\bibfield  {journal}
  {\bibinfo  {journal} {Science}\ }\textbf {\bibinfo {volume} {364}},\ \bibinfo
  {pages} {368} (\bibinfo {year} {2019})}\BibitemShut {NoStop}%
\bibitem [{\citenamefont {Magnard}\ \emph {et~al.}(2020)\citenamefont
  {Magnard}, \citenamefont {Storz}, \citenamefont {Kurpiers}, \citenamefont
  {Sch\"ar}, \citenamefont {Marxer}, \citenamefont {L\"utolf}, \citenamefont
  {Walter}, \citenamefont {Besse}, \citenamefont {Gabureac}, \citenamefont
  {Reuer}, \citenamefont {Akin}, \citenamefont {Royer}, \citenamefont {Blais},\
  and\ \citenamefont {Wallraff}}]{PhysRevLett.125.260502}%
  \BibitemOpen
  \bibfield  {author} {\bibinfo {author} {\bibfnamefont {P.}~\bibnamefont
  {Magnard}}, \bibinfo {author} {\bibfnamefont {S.}~\bibnamefont {Storz}},
  \bibinfo {author} {\bibfnamefont {P.}~\bibnamefont {Kurpiers}}, \bibinfo
  {author} {\bibfnamefont {J.}~\bibnamefont {Sch\"ar}}, \bibinfo {author}
  {\bibfnamefont {F.}~\bibnamefont {Marxer}}, \bibinfo {author} {\bibfnamefont
  {J.}~\bibnamefont {L\"utolf}}, \bibinfo {author} {\bibfnamefont
  {T.}~\bibnamefont {Walter}}, \bibinfo {author} {\bibfnamefont {J.-C.}\
  \bibnamefont {Besse}}, \bibinfo {author} {\bibfnamefont {M.}~\bibnamefont
  {Gabureac}}, \bibinfo {author} {\bibfnamefont {K.}~\bibnamefont {Reuer}},
  \bibinfo {author} {\bibfnamefont {A.}~\bibnamefont {Akin}}, \bibinfo {author}
  {\bibfnamefont {B.}~\bibnamefont {Royer}}, \bibinfo {author} {\bibfnamefont
  {A.}~\bibnamefont {Blais}},\ and\ \bibinfo {author} {\bibfnamefont
  {A.}~\bibnamefont {Wallraff}},\ }\bibfield  {title} {\bibinfo {title}
  {Microwave quantum link between superconducting circuits housed in spatially
  separated cryogenic systems},\ }\href
  {https://doi.org/10.1103/PhysRevLett.125.260502} {\bibfield  {journal}
  {\bibinfo  {journal} {Phys. Rev. Lett.}\ }\textbf {\bibinfo {volume} {125}},\
  \bibinfo {pages} {260502} (\bibinfo {year} {2020})}\BibitemShut {NoStop}%
\bibitem [{\citenamefont {Storz}\ \emph {et~al.}(2023)\citenamefont {Storz},
  \citenamefont {Sch{\"a}r}, \citenamefont {Kulikov}, \citenamefont {Magnard},
  \citenamefont {Kurpiers}, \citenamefont {L{\"u}tolf}, \citenamefont {Walter},
  \citenamefont {Copetudo}, \citenamefont {Reuer}, \citenamefont {Akin},
  \citenamefont {Besse}, \citenamefont {Gabureac}, \citenamefont {Norris},
  \citenamefont {Rosario}, \citenamefont {Martin}, \citenamefont {Martinez},
  \citenamefont {Amaya}, \citenamefont {Mitchell}, \citenamefont {Abellan},
  \citenamefont {Bancal}, \citenamefont {Sangouard}, \citenamefont {Royer},
  \citenamefont {Blais},\ and\ \citenamefont {Wallraff}}]{Storz2023}%
  \BibitemOpen
  \bibfield  {author} {\bibinfo {author} {\bibfnamefont {S.}~\bibnamefont
  {Storz}}, \bibinfo {author} {\bibfnamefont {J.}~\bibnamefont {Sch{\"a}r}},
  \bibinfo {author} {\bibfnamefont {A.}~\bibnamefont {Kulikov}}, \bibinfo
  {author} {\bibfnamefont {P.}~\bibnamefont {Magnard}}, \bibinfo {author}
  {\bibfnamefont {P.}~\bibnamefont {Kurpiers}}, \bibinfo {author}
  {\bibfnamefont {J.}~\bibnamefont {L{\"u}tolf}}, \bibinfo {author}
  {\bibfnamefont {T.}~\bibnamefont {Walter}}, \bibinfo {author} {\bibfnamefont
  {A.}~\bibnamefont {Copetudo}}, \bibinfo {author} {\bibfnamefont
  {K.}~\bibnamefont {Reuer}}, \bibinfo {author} {\bibfnamefont
  {A.}~\bibnamefont {Akin}}, \bibinfo {author} {\bibfnamefont {J.-C.}\
  \bibnamefont {Besse}}, \bibinfo {author} {\bibfnamefont {M.}~\bibnamefont
  {Gabureac}}, \bibinfo {author} {\bibfnamefont {G.~J.}\ \bibnamefont
  {Norris}}, \bibinfo {author} {\bibfnamefont {A.}~\bibnamefont {Rosario}},
  \bibinfo {author} {\bibfnamefont {F.}~\bibnamefont {Martin}}, \bibinfo
  {author} {\bibfnamefont {J.}~\bibnamefont {Martinez}}, \bibinfo {author}
  {\bibfnamefont {W.}~\bibnamefont {Amaya}}, \bibinfo {author} {\bibfnamefont
  {M.~W.}\ \bibnamefont {Mitchell}}, \bibinfo {author} {\bibfnamefont
  {C.}~\bibnamefont {Abellan}}, \bibinfo {author} {\bibfnamefont {J.-D.}\
  \bibnamefont {Bancal}}, \bibinfo {author} {\bibfnamefont {N.}~\bibnamefont
  {Sangouard}}, \bibinfo {author} {\bibfnamefont {B.}~\bibnamefont {Royer}},
  \bibinfo {author} {\bibfnamefont {A.}~\bibnamefont {Blais}},\ and\ \bibinfo
  {author} {\bibfnamefont {A.}~\bibnamefont {Wallraff}},\ }\bibfield  {title}
  {\bibinfo {title} {Loophole-free bell inequality violation with
  superconducting circuits},\ }\href
  {https://doi.org/10.1038/s41586-023-05885-0} {\bibfield  {journal} {\bibinfo
  {journal} {Nature}\ }\textbf {\bibinfo {volume} {617}},\ \bibinfo {pages}
  {265} (\bibinfo {year} {2023})}\BibitemShut {NoStop}%
\bibitem [{\citenamefont {Bennett}\ \emph {et~al.}(1993)\citenamefont
  {Bennett}, \citenamefont {Brassard}, \citenamefont {Cr\'epeau}, \citenamefont
  {Jozsa}, \citenamefont {Peres},\ and\ \citenamefont
  {Wootters}}]{PhysRevLett.70.1895}%
  \BibitemOpen
  \bibfield  {author} {\bibinfo {author} {\bibfnamefont {C.~H.}\ \bibnamefont
  {Bennett}}, \bibinfo {author} {\bibfnamefont {G.}~\bibnamefont {Brassard}},
  \bibinfo {author} {\bibfnamefont {C.}~\bibnamefont {Cr\'epeau}}, \bibinfo
  {author} {\bibfnamefont {R.}~\bibnamefont {Jozsa}}, \bibinfo {author}
  {\bibfnamefont {A.}~\bibnamefont {Peres}},\ and\ \bibinfo {author}
  {\bibfnamefont {W.~K.}\ \bibnamefont {Wootters}},\ }\bibfield  {title}
  {\bibinfo {title} {Teleporting an unknown quantum state via dual classical
  and einstein-podolsky-rosen channels},\ }\href
  {https://doi.org/10.1103/PhysRevLett.70.1895} {\bibfield  {journal} {\bibinfo
   {journal} {Phys. Rev. Lett.}\ }\textbf {\bibinfo {volume} {70}},\ \bibinfo
  {pages} {1895} (\bibinfo {year} {1993})}\BibitemShut {NoStop}%
\bibitem [{\citenamefont {\ifmmode~\dot{Z}\else \.{Z}\fi{}ukowski}\ \emph
  {et~al.}(1993)\citenamefont {\ifmmode~\dot{Z}\else \.{Z}\fi{}ukowski},
  \citenamefont {Zeilinger}, \citenamefont {Horne},\ and\ \citenamefont
  {Ekert}}]{PhysRevLett.71.4287}%
  \BibitemOpen
  \bibfield  {author} {\bibinfo {author} {\bibfnamefont {M.}~\bibnamefont
  {\ifmmode~\dot{Z}\else \.{Z}\fi{}ukowski}}, \bibinfo {author} {\bibfnamefont
  {A.}~\bibnamefont {Zeilinger}}, \bibinfo {author} {\bibfnamefont {M.~A.}\
  \bibnamefont {Horne}},\ and\ \bibinfo {author} {\bibfnamefont {A.~K.}\
  \bibnamefont {Ekert}},\ }\bibfield  {title} {\bibinfo {title}
  {``event-ready-detectors'' bell experiment via entanglement swapping},\
  }\href {https://doi.org/10.1103/PhysRevLett.71.4287} {\bibfield  {journal}
  {\bibinfo  {journal} {Phys. Rev. Lett.}\ }\textbf {\bibinfo {volume} {71}},\
  \bibinfo {pages} {4287} (\bibinfo {year} {1993})}\BibitemShut {NoStop}%
\bibitem [{\citenamefont {Pan}\ \emph {et~al.}(1998)\citenamefont {Pan},
  \citenamefont {Bouwmeester}, \citenamefont {Weinfurter},\ and\ \citenamefont
  {Zeilinger}}]{PhysRevLett.80.3891}%
  \BibitemOpen
  \bibfield  {author} {\bibinfo {author} {\bibfnamefont {J.-W.}\ \bibnamefont
  {Pan}}, \bibinfo {author} {\bibfnamefont {D.}~\bibnamefont {Bouwmeester}},
  \bibinfo {author} {\bibfnamefont {H.}~\bibnamefont {Weinfurter}},\ and\
  \bibinfo {author} {\bibfnamefont {A.}~\bibnamefont {Zeilinger}},\ }\bibfield
  {title} {\bibinfo {title} {Experimental entanglement swapping: Entangling
  photons that never interacted},\ }\href
  {https://doi.org/10.1103/PhysRevLett.80.3891} {\bibfield  {journal} {\bibinfo
   {journal} {Phys. Rev. Lett.}\ }\textbf {\bibinfo {volume} {80}},\ \bibinfo
  {pages} {3891} (\bibinfo {year} {1998})}\BibitemShut {NoStop}%
\bibitem [{\citenamefont {Briegel}\ \emph {et~al.}(1998)\citenamefont
  {Briegel}, \citenamefont {D\"ur}, \citenamefont {Cirac},\ and\ \citenamefont
  {Zoller}}]{PhysRevLett.81.5932}%
  \BibitemOpen
  \bibfield  {author} {\bibinfo {author} {\bibfnamefont {H.-J.}\ \bibnamefont
  {Briegel}}, \bibinfo {author} {\bibfnamefont {W.}~\bibnamefont {D\"ur}},
  \bibinfo {author} {\bibfnamefont {J.~I.}\ \bibnamefont {Cirac}},\ and\
  \bibinfo {author} {\bibfnamefont {P.}~\bibnamefont {Zoller}},\ }\bibfield
  {title} {\bibinfo {title} {Quantum repeaters: The role of imperfect local
  operations in quantum communication},\ }\href
  {https://doi.org/10.1103/PhysRevLett.81.5932} {\bibfield  {journal} {\bibinfo
   {journal} {Phys. Rev. Lett.}\ }\textbf {\bibinfo {volume} {81}},\ \bibinfo
  {pages} {5932} (\bibinfo {year} {1998})}\BibitemShut {NoStop}%
\bibitem [{\citenamefont {Duan}\ \emph {et~al.}(2001)\citenamefont {Duan},
  \citenamefont {Lukin}, \citenamefont {Cirac},\ and\ \citenamefont
  {Zoller}}]{Duan2001}%
  \BibitemOpen
  \bibfield  {author} {\bibinfo {author} {\bibfnamefont {L.-M.}\ \bibnamefont
  {Duan}}, \bibinfo {author} {\bibfnamefont {M.~D.}\ \bibnamefont {Lukin}},
  \bibinfo {author} {\bibfnamefont {J.~I.}\ \bibnamefont {Cirac}},\ and\
  \bibinfo {author} {\bibfnamefont {P.}~\bibnamefont {Zoller}},\ }\bibfield
  {title} {\bibinfo {title} {Long-distance quantum communication with atomic
  ensembles and linear optics},\ }\href {https://doi.org/10.1038/35106500}
  {\bibfield  {journal} {\bibinfo  {journal} {Nature}\ }\textbf {\bibinfo
  {volume} {414}},\ \bibinfo {pages} {413} (\bibinfo {year}
  {2001})}\BibitemShut {NoStop}%
\bibitem [{\citenamefont {Zhao}\ \emph {et~al.}(2003)\citenamefont {Zhao},
  \citenamefont {Yang}, \citenamefont {Chen}, \citenamefont {Zhang},\ and\
  \citenamefont {Pan}}]{PhysRevLett.90.207901}%
  \BibitemOpen
  \bibfield  {author} {\bibinfo {author} {\bibfnamefont {Z.}~\bibnamefont
  {Zhao}}, \bibinfo {author} {\bibfnamefont {T.}~\bibnamefont {Yang}}, \bibinfo
  {author} {\bibfnamefont {Y.-A.}\ \bibnamefont {Chen}}, \bibinfo {author}
  {\bibfnamefont {A.-N.}\ \bibnamefont {Zhang}},\ and\ \bibinfo {author}
  {\bibfnamefont {J.-W.}\ \bibnamefont {Pan}},\ }\bibfield  {title} {\bibinfo
  {title} {Experimental realization of entanglement concentration and a quantum
  repeater},\ }\href {https://doi.org/10.1103/PhysRevLett.90.207901} {\bibfield
   {journal} {\bibinfo  {journal} {Phys. Rev. Lett.}\ }\textbf {\bibinfo
  {volume} {90}},\ \bibinfo {pages} {207901} (\bibinfo {year}
  {2003})}\BibitemShut {NoStop}%
\bibitem [{\citenamefont {Sangouard}\ \emph {et~al.}(2011)\citenamefont
  {Sangouard}, \citenamefont {Simon}, \citenamefont {de~Riedmatten},\ and\
  \citenamefont {Gisin}}]{RevModPhys.83.33}%
  \BibitemOpen
  \bibfield  {author} {\bibinfo {author} {\bibfnamefont {N.}~\bibnamefont
  {Sangouard}}, \bibinfo {author} {\bibfnamefont {C.}~\bibnamefont {Simon}},
  \bibinfo {author} {\bibfnamefont {H.}~\bibnamefont {de~Riedmatten}},\ and\
  \bibinfo {author} {\bibfnamefont {N.}~\bibnamefont {Gisin}},\ }\bibfield
  {title} {\bibinfo {title} {Quantum repeaters based on atomic ensembles and
  linear optics},\ }\href {https://doi.org/10.1103/RevModPhys.83.33} {\bibfield
   {journal} {\bibinfo  {journal} {Rev. Mod. Phys.}\ }\textbf {\bibinfo
  {volume} {83}},\ \bibinfo {pages} {33} (\bibinfo {year} {2011})}\BibitemShut
  {NoStop}%
\bibitem [{\citenamefont {Azuma}\ \emph {et~al.}(2023)\citenamefont {Azuma},
  \citenamefont {Economou}, \citenamefont {Elkouss}, \citenamefont {Hilaire},
  \citenamefont {Jiang}, \citenamefont {Lo},\ and\ \citenamefont
  {Tzitrin}}]{RevModPhys.95.045006}%
  \BibitemOpen
  \bibfield  {author} {\bibinfo {author} {\bibfnamefont {K.}~\bibnamefont
  {Azuma}}, \bibinfo {author} {\bibfnamefont {S.~E.}\ \bibnamefont {Economou}},
  \bibinfo {author} {\bibfnamefont {D.}~\bibnamefont {Elkouss}}, \bibinfo
  {author} {\bibfnamefont {P.}~\bibnamefont {Hilaire}}, \bibinfo {author}
  {\bibfnamefont {L.}~\bibnamefont {Jiang}}, \bibinfo {author} {\bibfnamefont
  {H.-K.}\ \bibnamefont {Lo}},\ and\ \bibinfo {author} {\bibfnamefont
  {I.}~\bibnamefont {Tzitrin}},\ }\bibfield  {title} {\bibinfo {title} {Quantum
  repeaters: From quantum networks to the quantum internet},\ }\href
  {https://doi.org/10.1103/RevModPhys.95.045006} {\bibfield  {journal}
  {\bibinfo  {journal} {Rev. Mod. Phys.}\ }\textbf {\bibinfo {volume} {95}},\
  \bibinfo {pages} {045006} (\bibinfo {year} {2023})}\BibitemShut {NoStop}%
\bibitem [{\citenamefont {Ma}\ \emph {et~al.}(2012)\citenamefont {Ma},
  \citenamefont {Zotter}, \citenamefont {Kofler}, \citenamefont {Ursin},
  \citenamefont {Jennewein}, \citenamefont {Brukner},\ and\ \citenamefont
  {Zeilinger}}]{Ma2012}%
  \BibitemOpen
  \bibfield  {author} {\bibinfo {author} {\bibfnamefont {X.-s.}\ \bibnamefont
  {Ma}}, \bibinfo {author} {\bibfnamefont {S.}~\bibnamefont {Zotter}}, \bibinfo
  {author} {\bibfnamefont {J.}~\bibnamefont {Kofler}}, \bibinfo {author}
  {\bibfnamefont {R.}~\bibnamefont {Ursin}}, \bibinfo {author} {\bibfnamefont
  {T.}~\bibnamefont {Jennewein}}, \bibinfo {author} {\bibfnamefont
  {{\v{C}}.}~\bibnamefont {Brukner}},\ and\ \bibinfo {author} {\bibfnamefont
  {A.}~\bibnamefont {Zeilinger}},\ }\bibfield  {title} {\bibinfo {title}
  {Experimental delayed-choice entanglement swapping},\ }\href
  {https://doi.org/10.1038/nphys2294} {\bibfield  {journal} {\bibinfo
  {journal} {Nature Physics}\ }\textbf {\bibinfo {volume} {8}},\ \bibinfo
  {pages} {479} (\bibinfo {year} {2012})}\BibitemShut {NoStop}%
\bibitem [{\citenamefont {Hensen}\ \emph {et~al.}(2015)\citenamefont {Hensen},
  \citenamefont {Bernien}, \citenamefont {Dr{\'e}au}, \citenamefont {Reiserer},
  \citenamefont {Kalb}, \citenamefont {Blok}, \citenamefont {Ruitenberg},
  \citenamefont {Vermeulen}, \citenamefont {Schouten}, \citenamefont
  {Abell{\'a}n}, \citenamefont {Amaya}, \citenamefont {Pruneri}, \citenamefont
  {Mitchell}, \citenamefont {Markham}, \citenamefont {Twitchen}, \citenamefont
  {Elkouss}, \citenamefont {Wehner}, \citenamefont {Taminiau},\ and\
  \citenamefont {Hanson}}]{Hensen2015}%
  \BibitemOpen
  \bibfield  {author} {\bibinfo {author} {\bibfnamefont {B.}~\bibnamefont
  {Hensen}}, \bibinfo {author} {\bibfnamefont {H.}~\bibnamefont {Bernien}},
  \bibinfo {author} {\bibfnamefont {A.~E.}\ \bibnamefont {Dr{\'e}au}}, \bibinfo
  {author} {\bibfnamefont {A.}~\bibnamefont {Reiserer}}, \bibinfo {author}
  {\bibfnamefont {N.}~\bibnamefont {Kalb}}, \bibinfo {author} {\bibfnamefont
  {M.~S.}\ \bibnamefont {Blok}}, \bibinfo {author} {\bibfnamefont
  {J.}~\bibnamefont {Ruitenberg}}, \bibinfo {author} {\bibfnamefont {R.~F.~L.}\
  \bibnamefont {Vermeulen}}, \bibinfo {author} {\bibfnamefont {R.~N.}\
  \bibnamefont {Schouten}}, \bibinfo {author} {\bibfnamefont {C.}~\bibnamefont
  {Abell{\'a}n}}, \bibinfo {author} {\bibfnamefont {W.}~\bibnamefont {Amaya}},
  \bibinfo {author} {\bibfnamefont {V.}~\bibnamefont {Pruneri}}, \bibinfo
  {author} {\bibfnamefont {M.~W.}\ \bibnamefont {Mitchell}}, \bibinfo {author}
  {\bibfnamefont {M.}~\bibnamefont {Markham}}, \bibinfo {author} {\bibfnamefont
  {D.~J.}\ \bibnamefont {Twitchen}}, \bibinfo {author} {\bibfnamefont
  {D.}~\bibnamefont {Elkouss}}, \bibinfo {author} {\bibfnamefont
  {S.}~\bibnamefont {Wehner}}, \bibinfo {author} {\bibfnamefont {T.~H.}\
  \bibnamefont {Taminiau}},\ and\ \bibinfo {author} {\bibfnamefont
  {R.}~\bibnamefont {Hanson}},\ }\bibfield  {title} {\bibinfo {title}
  {Loophole-free bell inequality violation using electron spins separated by
  1.3 kilometres},\ }\href {https://doi.org/10.1038/nature15759} {\bibfield
  {journal} {\bibinfo  {journal} {Nature}\ }\textbf {\bibinfo {volume} {526}},\
  \bibinfo {pages} {682} (\bibinfo {year} {2015})}\BibitemShut {NoStop}%
\bibitem [{\citenamefont {Rosenfeld}\ \emph {et~al.}(2017)\citenamefont
  {Rosenfeld}, \citenamefont {Burchardt}, \citenamefont {Garthoff},
  \citenamefont {Redeker}, \citenamefont {Ortegel}, \citenamefont {Rau},\ and\
  \citenamefont {Weinfurter}}]{PhysRevLett.119.010402}%
  \BibitemOpen
  \bibfield  {author} {\bibinfo {author} {\bibfnamefont {W.}~\bibnamefont
  {Rosenfeld}}, \bibinfo {author} {\bibfnamefont {D.}~\bibnamefont
  {Burchardt}}, \bibinfo {author} {\bibfnamefont {R.}~\bibnamefont {Garthoff}},
  \bibinfo {author} {\bibfnamefont {K.}~\bibnamefont {Redeker}}, \bibinfo
  {author} {\bibfnamefont {N.}~\bibnamefont {Ortegel}}, \bibinfo {author}
  {\bibfnamefont {M.}~\bibnamefont {Rau}},\ and\ \bibinfo {author}
  {\bibfnamefont {H.}~\bibnamefont {Weinfurter}},\ }\bibfield  {title}
  {\bibinfo {title} {Event-ready bell test using entangled atoms simultaneously
  closing detection and locality loopholes},\ }\href
  {https://doi.org/10.1103/PhysRevLett.119.010402} {\bibfield  {journal}
  {\bibinfo  {journal} {Phys. Rev. Lett.}\ }\textbf {\bibinfo {volume} {119}},\
  \bibinfo {pages} {010402} (\bibinfo {year} {2017})}\BibitemShut {NoStop}%
\bibitem [{\citenamefont {Hong}\ \emph {et~al.}(1987)\citenamefont {Hong},
  \citenamefont {Ou},\ and\ \citenamefont {Mandel}}]{PhysRevLett.59.2044}%
  \BibitemOpen
  \bibfield  {author} {\bibinfo {author} {\bibfnamefont {C.~K.}\ \bibnamefont
  {Hong}}, \bibinfo {author} {\bibfnamefont {Z.~Y.}\ \bibnamefont {Ou}},\ and\
  \bibinfo {author} {\bibfnamefont {L.}~\bibnamefont {Mandel}},\ }\bibfield
  {title} {\bibinfo {title} {Measurement of subpicosecond time intervals
  between two photons by interference},\ }\href
  {https://doi.org/10.1103/PhysRevLett.59.2044} {\bibfield  {journal} {\bibinfo
   {journal} {Phys. Rev. Lett.}\ }\textbf {\bibinfo {volume} {59}},\ \bibinfo
  {pages} {2044} (\bibinfo {year} {1987})}\BibitemShut {NoStop}%
\bibitem [{\citenamefont {Riedmatten}\ \emph {et~al.}(2003)\citenamefont
  {Riedmatten}, \citenamefont {Marcikic}, \citenamefont {Tittel}, \citenamefont
  {Zbinden},\ and\ \citenamefont {Gisin}}]{PhysRevA.67.022301}%
  \BibitemOpen
  \bibfield  {author} {\bibinfo {author} {\bibfnamefont {H.}\ \bibnamefont
  {deRiedmatten}}, \bibinfo {author} {\bibfnamefont {I.}~\bibnamefont
  {Marcikic}}, \bibinfo {author} {\bibfnamefont {W.}~\bibnamefont {Tittel}},
  \bibinfo {author} {\bibfnamefont {H.}~\bibnamefont {Zbinden}},\ and\ \bibinfo
  {author} {\bibfnamefont {N.}~\bibnamefont {Gisin}},\ }\bibfield  {title}
  {\bibinfo {title} {Quantum interference with photon pairs created in
  spatially separated sources},\ }\href
  {https://doi.org/10.1103/PhysRevA.67.022301} {\bibfield  {journal} {\bibinfo
  {journal} {Phys. Rev. A}\ }\textbf {\bibinfo {volume} {67}},\ \bibinfo
  {pages} {022301} (\bibinfo {year} {2003})}\BibitemShut {NoStop}%
\bibitem [{\citenamefont {Yang}\ \emph {et~al.}(2006)\citenamefont {Yang},
  \citenamefont {Zhang}, \citenamefont {Chen}, \citenamefont {Lu},
  \citenamefont {Yin}, \citenamefont {Pan}, \citenamefont {Wei}, \citenamefont
  {Tian},\ and\ \citenamefont {Zhang}}]{PhysRevLett.96.110501}%
  \BibitemOpen
  \bibfield  {author} {\bibinfo {author} {\bibfnamefont {T.}~\bibnamefont
  {Yang}}, \bibinfo {author} {\bibfnamefont {Q.}~\bibnamefont {Zhang}},
  \bibinfo {author} {\bibfnamefont {T.-Y.}\ \bibnamefont {Chen}}, \bibinfo
  {author} {\bibfnamefont {S.}~\bibnamefont {Lu}}, \bibinfo {author}
  {\bibfnamefont {J.}~\bibnamefont {Yin}}, \bibinfo {author} {\bibfnamefont
  {J.-W.}\ \bibnamefont {Pan}}, \bibinfo {author} {\bibfnamefont {Z.-Y.}\
  \bibnamefont {Wei}}, \bibinfo {author} {\bibfnamefont {J.-R.}\ \bibnamefont
  {Tian}},\ and\ \bibinfo {author} {\bibfnamefont {J.}~\bibnamefont {Zhang}},\
  }\bibfield  {title} {\bibinfo {title} {Experimental synchronization of
  independent entangled photon sources},\ }\href
  {https://doi.org/10.1103/PhysRevLett.96.110501} {\bibfield  {journal}
  {\bibinfo  {journal} {Phys. Rev. Lett.}\ }\textbf {\bibinfo {volume} {96}},\
  \bibinfo {pages} {110501} (\bibinfo {year} {2006})}\BibitemShut {NoStop}%
\bibitem [{\citenamefont {Kaltenbaek}\ \emph {et~al.}(2006)\citenamefont
  {Kaltenbaek}, \citenamefont {Blauensteiner}, \citenamefont
  {\ifmmode~\dot{Z}\else \.{Z}\fi{}ukowski}, \citenamefont {Aspelmeyer},\ and\
  \citenamefont {Zeilinger}}]{PhysRevLett.96.240502}%
  \BibitemOpen
  \bibfield  {author} {\bibinfo {author} {\bibfnamefont {R.}~\bibnamefont
  {Kaltenbaek}}, \bibinfo {author} {\bibfnamefont {B.}~\bibnamefont
  {Blauensteiner}}, \bibinfo {author} {\bibfnamefont {M.}~\bibnamefont
  {\ifmmode~\dot{Z}\else \.{Z}\fi{}ukowski}}, \bibinfo {author} {\bibfnamefont
  {M.}~\bibnamefont {Aspelmeyer}},\ and\ \bibinfo {author} {\bibfnamefont
  {A.}~\bibnamefont {Zeilinger}},\ }\bibfield  {title} {\bibinfo {title}
  {Experimental interference of independent photons},\ }\href
  {https://doi.org/10.1103/PhysRevLett.96.240502} {\bibfield  {journal}
  {\bibinfo  {journal} {Phys. Rev. Lett.}\ }\textbf {\bibinfo {volume} {96}},\
  \bibinfo {pages} {240502} (\bibinfo {year} {2006})}\BibitemShut {NoStop}%
\bibitem [{\citenamefont {Kaltenbaek}\ \emph {et~al.}(2009)\citenamefont
  {Kaltenbaek}, \citenamefont {Prevedel}, \citenamefont {Aspelmeyer},\ and\
  \citenamefont {Zeilinger}}]{PhysRevA.79.040302}%
  \BibitemOpen
  \bibfield  {author} {\bibinfo {author} {\bibfnamefont {R.}~\bibnamefont
  {Kaltenbaek}}, \bibinfo {author} {\bibfnamefont {R.}~\bibnamefont
  {Prevedel}}, \bibinfo {author} {\bibfnamefont {M.}~\bibnamefont
  {Aspelmeyer}},\ and\ \bibinfo {author} {\bibfnamefont {A.}~\bibnamefont
  {Zeilinger}},\ }\bibfield  {title} {\bibinfo {title} {High-fidelity
  entanglement swapping with fully independent sources},\ }\href
  {https://doi.org/10.1103/PhysRevA.79.040302} {\bibfield  {journal} {\bibinfo
  {journal} {Phys. Rev. A}\ }\textbf {\bibinfo {volume} {79}},\ \bibinfo
  {pages} {040302(R)} (\bibinfo {year} {2009})}\BibitemShut {NoStop}%
\bibitem [{\citenamefont {Zou}\ \emph {et~al.}(1991)\citenamefont {Zou},
  \citenamefont {Wang},\ and\ \citenamefont {Mandel}}]{PhysRevLett.67.318}%
  \BibitemOpen
  \bibfield  {author} {\bibinfo {author} {\bibfnamefont {X.~Y.}\ \bibnamefont
  {Zou}}, \bibinfo {author} {\bibfnamefont {L.~J.}\ \bibnamefont {Wang}},\ and\
  \bibinfo {author} {\bibfnamefont {L.}~\bibnamefont {Mandel}},\ }\bibfield
  {title} {\bibinfo {title} {Induced coherence and indistinguishability in
  optical interference},\ }\href {https://doi.org/10.1103/PhysRevLett.67.318}
  {\bibfield  {journal} {\bibinfo  {journal} {Phys. Rev. Lett.}\ }\textbf
  {\bibinfo {volume} {67}},\ \bibinfo {pages} {318} (\bibinfo {year}
  {1991})}\BibitemShut {NoStop}%
\bibitem [{\citenamefont {Herzog}\ \emph {et~al.}(1994)\citenamefont {Herzog},
  \citenamefont {Rarity}, \citenamefont {Weinfurter},\ and\ \citenamefont
  {Zeilinger}}]{PhysRevLett.72.629}%
  \BibitemOpen
  \bibfield  {author} {\bibinfo {author} {\bibfnamefont {T.~J.}\ \bibnamefont
  {Herzog}}, \bibinfo {author} {\bibfnamefont {J.~G.}\ \bibnamefont {Rarity}},
  \bibinfo {author} {\bibfnamefont {H.}~\bibnamefont {Weinfurter}},\ and\
  \bibinfo {author} {\bibfnamefont {A.}~\bibnamefont {Zeilinger}},\ }\bibfield
  {title} {\bibinfo {title} {Frustrated two-photon creation via interference},\
  }\href {https://doi.org/10.1103/PhysRevLett.72.629} {\bibfield  {journal}
  {\bibinfo  {journal} {Phys. Rev. Lett.}\ }\textbf {\bibinfo {volume} {72}},\
  \bibinfo {pages} {629} (\bibinfo {year} {1994})}\BibitemShut {NoStop}%
\bibitem [{\citenamefont {Greenberger}\ \emph {et~al.}(1993)\citenamefont
  {Greenberger}, \citenamefont {Horne},\ and\ \citenamefont
  {Zeilinger}}]{10.1063/1.881360}%
  \BibitemOpen
  \bibfield  {author} {\bibinfo {author} {\bibfnamefont {D.~M.}\ \bibnamefont
  {Greenberger}}, \bibinfo {author} {\bibfnamefont {M.~A.}\ \bibnamefont
  {Horne}},\ and\ \bibinfo {author} {\bibfnamefont {A.}~\bibnamefont
  {Zeilinger}},\ }\bibfield  {title} {\bibinfo {title} {{Multiparticle
  Interferometry and the Superposition Principle}},\ }\href
  {https://doi.org/10.1063/1.881360} {\bibfield  {journal} {\bibinfo  {journal}
  {Physics Today}\ }\textbf {\bibinfo {volume} {46}},\ \bibinfo {pages} {22}
  (\bibinfo {year} {1993})}\BibitemShut {NoStop}%
\bibitem [{\citenamefont {Hochrainer}\ \emph {et~al.}(2022)\citenamefont
  {Hochrainer}, \citenamefont {Lahiri}, \citenamefont {Erhard}, \citenamefont
  {Krenn},\ and\ \citenamefont {Zeilinger}}]{RevModPhys.94.025007}%
  \BibitemOpen
  \bibfield  {author} {\bibinfo {author} {\bibfnamefont {A.}~\bibnamefont
  {Hochrainer}}, \bibinfo {author} {\bibfnamefont {M.}~\bibnamefont {Lahiri}},
  \bibinfo {author} {\bibfnamefont {M.}~\bibnamefont {Erhard}}, \bibinfo
  {author} {\bibfnamefont {M.}~\bibnamefont {Krenn}},\ and\ \bibinfo {author}
  {\bibfnamefont {A.}~\bibnamefont {Zeilinger}},\ }\bibfield  {title} {\bibinfo
  {title} {Quantum indistinguishability by path identity and with undetected
  photons},\ }\href {https://doi.org/10.1103/RevModPhys.94.025007} {\bibfield
  {journal} {\bibinfo  {journal} {Rev. Mod. Phys.}\ }\textbf {\bibinfo {volume}
  {94}},\ \bibinfo {pages} {025007} (\bibinfo {year} {2022})}\BibitemShut
  {NoStop}%
\bibitem [{\citenamefont {Ruiz-Gonzalez}\ \emph {et~al.}(2023)\citenamefont
  {Ruiz-Gonzalez}, \citenamefont {Arlt}, \citenamefont {Petermann},
  \citenamefont {Sayyad}, \citenamefont {Jaouni}, \citenamefont {Karimi},
  \citenamefont {Tischler}, \citenamefont {Gu},\ and\ \citenamefont
  {Krenn}}]{RuizGonzalez2023digitaldiscoveryof}%
  \BibitemOpen
  \bibfield  {author} {\bibinfo {author} {\bibfnamefont {C.}~\bibnamefont
  {Ruiz-Gonzalez}}, \bibinfo {author} {\bibfnamefont {S.}~\bibnamefont {Arlt}},
  \bibinfo {author} {\bibfnamefont {J.}~\bibnamefont {Petermann}}, \bibinfo
  {author} {\bibfnamefont {S.}~\bibnamefont {Sayyad}}, \bibinfo {author}
  {\bibfnamefont {T.}~\bibnamefont {Jaouni}}, \bibinfo {author} {\bibfnamefont
  {E.}~\bibnamefont {Karimi}}, \bibinfo {author} {\bibfnamefont
  {N.}~\bibnamefont {Tischler}}, \bibinfo {author} {\bibfnamefont
  {X.}~\bibnamefont {Gu}},\ and\ \bibinfo {author} {\bibfnamefont
  {M.}~\bibnamefont {Krenn}},\ }\bibfield  {title} {\bibinfo {title} {Digital
  {D}iscovery of 100 diverse {Q}uantum {E}xperiments with {P}y{T}heus},\ }\href
  {https://doi.org/10.22331/q-2023-12-12-1204} {\bibfield  {journal} {\bibinfo
  {journal} {{Quantum}}\ }\textbf {\bibinfo {volume} {7}},\ \bibinfo {pages}
  {1204} (\bibinfo {year} {2023})}\BibitemShut {NoStop}%
\bibitem [{\citenamefont {Gu}\ \emph {et~al.}(2019)\citenamefont {Gu},
  \citenamefont {Erhard}, \citenamefont {Zeilinger},\ and\ \citenamefont
  {Krenn}}]{Gu2019}%
  \BibitemOpen
  \bibfield  {author} {\bibinfo {author} {\bibfnamefont {X.}~\bibnamefont
  {Gu}}, \bibinfo {author} {\bibfnamefont {M.}~\bibnamefont {Erhard}}, \bibinfo
  {author} {\bibfnamefont {A.}~\bibnamefont {Zeilinger}},\ and\ \bibinfo
  {author} {\bibfnamefont {M.}~\bibnamefont {Krenn}},\ }\bibfield  {title}
  {\bibinfo {title} {Quantum experiments and graphs ii: Quantum interference,
  computation, and state generation},\ }\href
  {https://doi.org/10.1073/pnas.1815884116} {\bibfield  {journal} {\bibinfo
  {journal} {Proceedings of the National Academy of Sciences}\ }\textbf
  {\bibinfo {volume} {116}},\ \bibinfo {pages} {4147} (\bibinfo {year}
  {2019})}\BibitemShut {NoStop}%
\bibitem [{\citenamefont {Feng}\ \emph {et~al.}(2023)\citenamefont {Feng},
  \citenamefont {Zhang}, \citenamefont {Liu}, \citenamefont {Cheng},
  \citenamefont {Guo}, \citenamefont {Dai}, \citenamefont {Guo}, \citenamefont
  {Krenn},\ and\ \citenamefont {Ren}}]{Feng:23}%
  \BibitemOpen
  \bibfield  {author} {\bibinfo {author} {\bibfnamefont {L.-T.}\ \bibnamefont
  {Feng}}, \bibinfo {author} {\bibfnamefont {M.}~\bibnamefont {Zhang}},
  \bibinfo {author} {\bibfnamefont {D.}~\bibnamefont {Liu}}, \bibinfo {author}
  {\bibfnamefont {Y.-J.}\ \bibnamefont {Cheng}}, \bibinfo {author}
  {\bibfnamefont {G.-P.}\ \bibnamefont {Guo}}, \bibinfo {author} {\bibfnamefont
  {D.-X.}\ \bibnamefont {Dai}}, \bibinfo {author} {\bibfnamefont {G.-C.}\
  \bibnamefont {Guo}}, \bibinfo {author} {\bibfnamefont {M.}~\bibnamefont
  {Krenn}},\ and\ \bibinfo {author} {\bibfnamefont {X.-F.}\ \bibnamefont
  {Ren}},\ }\bibfield  {title} {\bibinfo {title} {On-chip quantum interference
  between the origins of a multi-photon state},\ }\href
  {https://doi.org/10.1364/OPTICA.474750} {\bibfield  {journal} {\bibinfo
  {journal} {Optica}\ }\textbf {\bibinfo {volume} {10}},\ \bibinfo {pages}
  {105} (\bibinfo {year} {2023})}\BibitemShut {NoStop}%
\bibitem [{\citenamefont {Qian}\ \emph {et~al.}(2023)\citenamefont {Qian},
  \citenamefont {Wang}, \citenamefont {Chen}, \citenamefont {Hou},
  \citenamefont {Krenn}, \citenamefont {Zhu},\ and\ \citenamefont
  {Ma}}]{Qian2023}%
  \BibitemOpen
  \bibfield  {author} {\bibinfo {author} {\bibfnamefont {K.}~\bibnamefont
  {Qian}}, \bibinfo {author} {\bibfnamefont {K.}~\bibnamefont {Wang}}, \bibinfo
  {author} {\bibfnamefont {L.}~\bibnamefont {Chen}}, \bibinfo {author}
  {\bibfnamefont {Z.}~\bibnamefont {Hou}}, \bibinfo {author} {\bibfnamefont
  {M.}~\bibnamefont {Krenn}}, \bibinfo {author} {\bibfnamefont
  {S.}~\bibnamefont {Zhu}},\ and\ \bibinfo {author} {\bibfnamefont {X.-s.}\
  \bibnamefont {Ma}},\ }\bibfield  {title} {\bibinfo {title} {Multiphoton
  non-local quantum interference controlled by an undetected photon},\ }\href
  {https://doi.org/10.1038/s41467-023-37228-y} {\bibfield  {journal} {\bibinfo
  {journal} {Nature Communications}\ }\textbf {\bibinfo {volume} {14}},\
  \bibinfo {pages} {1480} (\bibinfo {year} {2023})}\BibitemShut {NoStop}%
\bibitem [{\citenamefont {Resch}\ \emph {et~al.}(2001)\citenamefont {Resch},
  \citenamefont {Lundeen},\ and\ \citenamefont
  {Steinberg}}]{PhysRevLett.87.123603}%
  \BibitemOpen
  \bibfield  {author} {\bibinfo {author} {\bibfnamefont {K.~J.}\ \bibnamefont
  {Resch}}, \bibinfo {author} {\bibfnamefont {J.~S.}\ \bibnamefont {Lundeen}},\
  and\ \bibinfo {author} {\bibfnamefont {A.~M.}\ \bibnamefont {Steinberg}},\
  }\bibfield  {title} {\bibinfo {title} {Nonlinear optics with less than one
  photon},\ }\href {https://doi.org/10.1103/PhysRevLett.87.123603} {\bibfield
  {journal} {\bibinfo  {journal} {Phys. Rev. Lett.}\ }\textbf {\bibinfo
  {volume} {87}},\ \bibinfo {pages} {123603} (\bibinfo {year}
  {2001})}\BibitemShut {NoStop}%
\bibitem [{\citenamefont {Takeuchi}(2001)}]{Takeuchi:01}%
  \BibitemOpen
  \bibfield  {author} {\bibinfo {author} {\bibfnamefont {S.}~\bibnamefont
  {Takeuchi}},\ }\bibfield  {title} {\bibinfo {title} {Beamlike twin-photon
  generation by use of type ii parametric downconversion},\ }\href
  {https://doi.org/10.1364/OL.26.000843} {\bibfield  {journal} {\bibinfo
  {journal} {Opt. Lett.}\ }\textbf {\bibinfo {volume} {26}},\ \bibinfo {pages}
  {843} (\bibinfo {year} {2001})}\BibitemShut {NoStop}%
\bibitem [{\citenamefont {Niu}\ \emph {et~al.}(2008)\citenamefont {Niu},
  \citenamefont {Huang}, \citenamefont {Xiang}, \citenamefont {Guo},\ and\
  \citenamefont {Ou}}]{Niu:08}%
  \BibitemOpen
  \bibfield  {author} {\bibinfo {author} {\bibfnamefont {X.-L.}\ \bibnamefont
  {Niu}}, \bibinfo {author} {\bibfnamefont {Y.-F.}\ \bibnamefont {Huang}},
  \bibinfo {author} {\bibfnamefont {G.-Y.}\ \bibnamefont {Xiang}}, \bibinfo
  {author} {\bibfnamefont {G.-C.}\ \bibnamefont {Guo}},\ and\ \bibinfo {author}
  {\bibfnamefont {Z.~Y.}\ \bibnamefont {Ou}},\ }\bibfield  {title} {\bibinfo
  {title} {Beamlike high-brightness source of polarization-entangled photon
  pairs},\ }\href {https://doi.org/10.1364/OL.33.000968} {\bibfield  {journal}
  {\bibinfo  {journal} {Opt. Lett.}\ }\textbf {\bibinfo {volume} {33}},\
  \bibinfo {pages} {968} (\bibinfo {year} {2008})}\BibitemShut {NoStop}%
\bibitem [{\citenamefont {Clauser}\ \emph {et~al.}(1969)\citenamefont
  {Clauser}, \citenamefont {Horne}, \citenamefont {Shimony},\ and\
  \citenamefont {Holt}}]{PhysRevLett.23.880}%
  \BibitemOpen
  \bibfield  {author} {\bibinfo {author} {\bibfnamefont {J.~F.}\ \bibnamefont
  {Clauser}}, \bibinfo {author} {\bibfnamefont {M.~A.}\ \bibnamefont {Horne}},
  \bibinfo {author} {\bibfnamefont {A.}~\bibnamefont {Shimony}},\ and\ \bibinfo
  {author} {\bibfnamefont {R.~A.}\ \bibnamefont {Holt}},\ }\bibfield  {title}
  {\bibinfo {title} {Proposed experiment to test local hidden-variable
  theories},\ }\href {https://doi.org/10.1103/PhysRevLett.23.880} {\bibfield
  {journal} {\bibinfo  {journal} {Phys. Rev. Lett.}\ }\textbf {\bibinfo
  {volume} {23}},\ \bibinfo {pages} {880} (\bibinfo {year} {1969})}\BibitemShut
  {NoStop}%
\bibitem [{\citenamefont {Peres}(1996)}]{PhysRevLett.77.1413}%
  \BibitemOpen
  \bibfield  {author} {\bibinfo {author} {\bibfnamefont {A.}~\bibnamefont
  {Peres}},\ }\bibfield  {title} {\bibinfo {title} {Separability criterion for
  density matrices},\ }\href {https://doi.org/10.1103/PhysRevLett.77.1413}
  {\bibfield  {journal} {\bibinfo  {journal} {Phys. Rev. Lett.}\ }\textbf
  {\bibinfo {volume} {77}},\ \bibinfo {pages} {1413} (\bibinfo {year}
  {1996})}\BibitemShut {NoStop}%
\bibitem [{\citenamefont {Bourennane}\ \emph {et~al.}(2004)\citenamefont
  {Bourennane}, \citenamefont {Eibl}, \citenamefont {Kurtsiefer}, \citenamefont
  {Gaertner}, \citenamefont {Weinfurter}, \citenamefont {G\"uhne},
  \citenamefont {Hyllus}, \citenamefont {Bru\ss{}}, \citenamefont
  {Lewenstein},\ and\ \citenamefont {Sanpera}}]{PhysRevLett.92.087902}%
  \BibitemOpen
  \bibfield  {author} {\bibinfo {author} {\bibfnamefont {M.}~\bibnamefont
  {Bourennane}}, \bibinfo {author} {\bibfnamefont {M.}~\bibnamefont {Eibl}},
  \bibinfo {author} {\bibfnamefont {C.}~\bibnamefont {Kurtsiefer}}, \bibinfo
  {author} {\bibfnamefont {S.}~\bibnamefont {Gaertner}}, \bibinfo {author}
  {\bibfnamefont {H.}~\bibnamefont {Weinfurter}}, \bibinfo {author}
  {\bibfnamefont {O.}~\bibnamefont {G\"uhne}}, \bibinfo {author} {\bibfnamefont
  {P.}~\bibnamefont {Hyllus}}, \bibinfo {author} {\bibfnamefont
  {D.}~\bibnamefont {Bru\ss{}}}, \bibinfo {author} {\bibfnamefont
  {M.}~\bibnamefont {Lewenstein}},\ and\ \bibinfo {author} {\bibfnamefont
  {A.}~\bibnamefont {Sanpera}},\ }\bibfield  {title} {\bibinfo {title}
  {Experimental detection of multipartite entanglement using witness
  operators},\ }\href {https://doi.org/10.1103/PhysRevLett.92.087902}
  {\bibfield  {journal} {\bibinfo  {journal} {Phys. Rev. Lett.}\ }\textbf
  {\bibinfo {volume} {92}},\ \bibinfo {pages} {087902} (\bibinfo {year}
  {2004})}\BibitemShut {NoStop}%
\bibitem [{\citenamefont {Wang}\ \emph {et~al.}(2015)\citenamefont {Wang},
  \citenamefont {Cai}, \citenamefont {Su}, \citenamefont {Chen}, \citenamefont
  {Wu}, \citenamefont {Li}, \citenamefont {Liu}, \citenamefont {Lu},\ and\
  \citenamefont {Pan}}]{Wang2015}%
  \BibitemOpen
  \bibfield  {author} {\bibinfo {author} {\bibfnamefont {X.-L.}\ \bibnamefont
  {Wang}}, \bibinfo {author} {\bibfnamefont {X.-D.}\ \bibnamefont {Cai}},
  \bibinfo {author} {\bibfnamefont {Z.-E.}\ \bibnamefont {Su}}, \bibinfo
  {author} {\bibfnamefont {M.-C.}\ \bibnamefont {Chen}}, \bibinfo {author}
  {\bibfnamefont {D.}~\bibnamefont {Wu}}, \bibinfo {author} {\bibfnamefont
  {L.}~\bibnamefont {Li}}, \bibinfo {author} {\bibfnamefont {N.-L.}\
  \bibnamefont {Liu}}, \bibinfo {author} {\bibfnamefont {C.-Y.}\ \bibnamefont
  {Lu}},\ and\ \bibinfo {author} {\bibfnamefont {J.-W.}\ \bibnamefont {Pan}},\
  }\bibfield  {title} {\bibinfo {title} {Quantum teleportation of multiple
  degrees of freedom of a single photon},\ }\href
  {https://doi.org/10.1038/nature14246} {\bibfield  {journal} {\bibinfo
  {journal} {Nature}\ }\textbf {\bibinfo {volume} {518}},\ \bibinfo {pages}
  {516} (\bibinfo {year} {2015})}\BibitemShut {NoStop}%
\bibitem [{\citenamefont {Luo}\ \emph {et~al.}(2019)\citenamefont {Luo},
  \citenamefont {Zhong}, \citenamefont {Erhard}, \citenamefont {Wang},
  \citenamefont {Peng}, \citenamefont {Krenn}, \citenamefont {Jiang},
  \citenamefont {Li}, \citenamefont {Liu}, \citenamefont {Lu}, \citenamefont
  {Zeilinger},\ and\ \citenamefont {Pan}}]{PhysRevLett.123.070505}%
  \BibitemOpen
  \bibfield  {author} {\bibinfo {author} {\bibfnamefont {Y.-H.}\ \bibnamefont
  {Luo}}, \bibinfo {author} {\bibfnamefont {H.-S.}\ \bibnamefont {Zhong}},
  \bibinfo {author} {\bibfnamefont {M.}~\bibnamefont {Erhard}}, \bibinfo
  {author} {\bibfnamefont {X.-L.}\ \bibnamefont {Wang}}, \bibinfo {author}
  {\bibfnamefont {L.-C.}\ \bibnamefont {Peng}}, \bibinfo {author}
  {\bibfnamefont {M.}~\bibnamefont {Krenn}}, \bibinfo {author} {\bibfnamefont
  {X.}~\bibnamefont {Jiang}}, \bibinfo {author} {\bibfnamefont
  {L.}~\bibnamefont {Li}}, \bibinfo {author} {\bibfnamefont {N.-L.}\
  \bibnamefont {Liu}}, \bibinfo {author} {\bibfnamefont {C.-Y.}\ \bibnamefont
  {Lu}}, \bibinfo {author} {\bibfnamefont {A.}~\bibnamefont {Zeilinger}},\ and\
  \bibinfo {author} {\bibfnamefont {J.-W.}\ \bibnamefont {Pan}},\ }\bibfield
  {title} {\bibinfo {title} {Quantum teleportation in high dimensions},\ }\href
  {https://doi.org/10.1103/PhysRevLett.123.070505} {\bibfield  {journal}
  {\bibinfo  {journal} {Phys. Rev. Lett.}\ }\textbf {\bibinfo {volume} {123}},\
  \bibinfo {pages} {070505} (\bibinfo {year} {2019})}\BibitemShut {NoStop}%
\bibitem [{\citenamefont {Hu}\ \emph {et~al.}(2020{\natexlab{a}})\citenamefont
  {Hu}, \citenamefont {Zhang}, \citenamefont {Liu}, \citenamefont {Cai},
  \citenamefont {Ye}, \citenamefont {Guo}, \citenamefont {Xing}, \citenamefont
  {Huang}, \citenamefont {Huang}, \citenamefont {Li},\ and\ \citenamefont
  {Guo}}]{PhysRevLett.125.230501}%
  \BibitemOpen
  \bibfield  {author} {\bibinfo {author} {\bibfnamefont {X.-M.}\ \bibnamefont
  {Hu}}, \bibinfo {author} {\bibfnamefont {C.}~\bibnamefont {Zhang}}, \bibinfo
  {author} {\bibfnamefont {B.-H.}\ \bibnamefont {Liu}}, \bibinfo {author}
  {\bibfnamefont {Y.}~\bibnamefont {Cai}}, \bibinfo {author} {\bibfnamefont
  {X.-J.}\ \bibnamefont {Ye}}, \bibinfo {author} {\bibfnamefont
  {Y.}~\bibnamefont {Guo}}, \bibinfo {author} {\bibfnamefont {W.-B.}\
  \bibnamefont {Xing}}, \bibinfo {author} {\bibfnamefont {C.-X.}\ \bibnamefont
  {Huang}}, \bibinfo {author} {\bibfnamefont {Y.-F.}\ \bibnamefont {Huang}},
  \bibinfo {author} {\bibfnamefont {C.-F.}\ \bibnamefont {Li}},\ and\ \bibinfo
  {author} {\bibfnamefont {G.-C.}\ \bibnamefont {Guo}},\ }\bibfield  {title}
  {\bibinfo {title} {Experimental high-dimensional quantum teleportation},\
  }\href {https://doi.org/10.1103/PhysRevLett.125.230501} {\bibfield  {journal}
  {\bibinfo  {journal} {Phys. Rev. Lett.}\ }\textbf {\bibinfo {volume} {125}},\
  \bibinfo {pages} {230501} (\bibinfo {year} {2020}{\natexlab{a}})}\BibitemShut
  {NoStop}%
\bibitem [{\citenamefont {Pivoluska}\ \emph {et~al.}(2018)\citenamefont
  {Pivoluska}, \citenamefont {Huber},\ and\ \citenamefont
  {Malik}}]{PhysRevA.97.032312}%
  \BibitemOpen
  \bibfield  {author} {\bibinfo {author} {\bibfnamefont {M.}~\bibnamefont
  {Pivoluska}}, \bibinfo {author} {\bibfnamefont {M.}~\bibnamefont {Huber}},\
  and\ \bibinfo {author} {\bibfnamefont {M.}~\bibnamefont {Malik}},\ }\bibfield
   {title} {\bibinfo {title} {Layered quantum key distribution},\ }\href
  {https://doi.org/10.1103/PhysRevA.97.032312} {\bibfield  {journal} {\bibinfo
  {journal} {Phys. Rev. A}\ }\textbf {\bibinfo {volume} {97}},\ \bibinfo
  {pages} {032312} (\bibinfo {year} {2018})}\BibitemShut {NoStop}%
\bibitem [{\citenamefont {Hu}\ \emph {et~al.}(2020{\natexlab{b}})\citenamefont
  {Hu}, \citenamefont {Xing}, \citenamefont {Zhang}, \citenamefont {Liu},
  \citenamefont {Pivoluska}, \citenamefont {Huber}, \citenamefont {Huang},
  \citenamefont {Li},\ and\ \citenamefont {Guo}}]{Hu2020}%
  \BibitemOpen
  \bibfield  {author} {\bibinfo {author} {\bibfnamefont {X.-M.}\ \bibnamefont
  {Hu}}, \bibinfo {author} {\bibfnamefont {W.-B.}\ \bibnamefont {Xing}},
  \bibinfo {author} {\bibfnamefont {C.}~\bibnamefont {Zhang}}, \bibinfo
  {author} {\bibfnamefont {B.-H.}\ \bibnamefont {Liu}}, \bibinfo {author}
  {\bibfnamefont {M.}~\bibnamefont {Pivoluska}}, \bibinfo {author}
  {\bibfnamefont {M.}~\bibnamefont {Huber}}, \bibinfo {author} {\bibfnamefont
  {Y.-F.}\ \bibnamefont {Huang}}, \bibinfo {author} {\bibfnamefont {C.-F.}\
  \bibnamefont {Li}},\ and\ \bibinfo {author} {\bibfnamefont {G.-C.}\
  \bibnamefont {Guo}},\ }\bibfield  {title} {\bibinfo {title} {Experimental
  creation of multi-photon high-dimensional layered quantum states},\ }\href
  {https://doi.org/10.1038/s41534-020-00318-6} {\bibfield  {journal} {\bibinfo
  {journal} {npj Quantum Information}\ }\textbf {\bibinfo {volume} {6}},\
  \bibinfo {pages} {88} (\bibinfo {year} {2020}{\natexlab{b}})}\BibitemShut
  {NoStop}%
\bibitem [{\citenamefont {Wang}\ \emph {et~al.}(2022)\citenamefont {Wang},
  \citenamefont {Hao}, \citenamefont {Liu}, \citenamefont {Sun}, \citenamefont
  {Xu}, \citenamefont {Li}, \citenamefont {Guo}, \citenamefont {Castellini},
  \citenamefont {Bellomo}, \citenamefont {Compagno},\ and\ \citenamefont
  {Lo~Franco}}]{PhysRevA.106.032609}%
  \BibitemOpen
  \bibfield  {author} {\bibinfo {author} {\bibfnamefont {Y.}~\bibnamefont
  {Wang}}, \bibinfo {author} {\bibfnamefont {Z.-Y.}\ \bibnamefont {Hao}},
  \bibinfo {author} {\bibfnamefont {Z.-H.}\ \bibnamefont {Liu}}, \bibinfo
  {author} {\bibfnamefont {K.}~\bibnamefont {Sun}}, \bibinfo {author}
  {\bibfnamefont {J.-S.}\ \bibnamefont {Xu}}, \bibinfo {author} {\bibfnamefont
  {C.-F.}\ \bibnamefont {Li}}, \bibinfo {author} {\bibfnamefont {G.-C.}\
  \bibnamefont {Guo}}, \bibinfo {author} {\bibfnamefont {A.}~\bibnamefont
  {Castellini}}, \bibinfo {author} {\bibfnamefont {B.}~\bibnamefont {Bellomo}},
  \bibinfo {author} {\bibfnamefont {G.}~\bibnamefont {Compagno}},\ and\
  \bibinfo {author} {\bibfnamefont {R.}~\bibnamefont {Lo~Franco}},\ }\bibfield
  {title} {\bibinfo {title} {Remote entanglement distribution in a quantum
  network via multinode indistinguishability of photons},\ }\href
  {https://doi.org/10.1103/PhysRevA.106.032609} {\bibfield  {journal} {\bibinfo
   {journal} {Phys. Rev. A}\ }\textbf {\bibinfo {volume} {106}},\ \bibinfo
  {pages} {032609} (\bibinfo {year} {2022})}\BibitemShut {NoStop}%
\end{thebibliography}
%

\end{document}